\documentclass[12pt]{article}

\usepackage{graphicx,stfloats}

\usepackage{amsthm}
\usepackage{algorithmic}
\usepackage{algorithm}
\usepackage{bm}
\usepackage{amsmath,amssymb}

\theoremstyle{plain}
\theoremstyle{remark}
\newtheorem{remark}{Remark}

\begin{document}
\begin{center}
	
	{\fontsize{15pt}{5pt}\selectfont
	Formula of Laminar Flame Speed Coupled with Differential Equation for Temperature in Low-Mach-Number Model
	}\\ 
	\vspace{10pt}
	
	{\fontsize{12pt}{5pt}\selectfont Keigo Wada}\footnote{\fontsize{12pt}{5pt}\selectfont Institute of Philosophy in Interdisciplinary Sciences, Kanazawa University, Kakuma, Kanazawa, Ishikawa 920-1192, Japan}
	\footnote{\fontsize{12pt}{5pt}\selectfont Additional Post Member of OCAMI, Osaka Metropolitan University, Sugimoto, Sumiyoshi-ku, Osaka 558-8585, Japan}
	
	{\fontsize{12pt}{5pt}\selectfont k-wada@kyudai.jp}\\
	
	\vspace{10pt}

	\begin{abstract}
		The non-monotonic profile of temperature is to be considered in the context of combustion inside tubes or thermonuclear flames, which may accelerates to become detonation waves. This transition is known as deflagration-to-detonation transition.
		Focusing on this point, we extend the study of adiabatic flames, which assumes zero temperature gradient at the burned-side edge of a flame front, to the case of non-zero temperature gradient by introducing compression zone behind the front.
		Such a treatment has been performed in the previous work based on the ignition temperature approximation.
		In that case, the effect of gas compression on flames was described by asymptotic solutions with respect to small squared Mach numbers: for example, in the compression zone, inner solutions are expressed by use of Lambert $W$ function.
		However, due to the aim of obtaining analytical solutions inside a flame front, the temperature dependence of reaction term was ignored and, as a result, the cold boundary difficulty was not treated seriously.
		This issue is resolved in this study by use of large activation energy asymptotics with the reaction term expressed by the Arrhenius type.
		The temperature distribution is computed numerically in the reaction zone and calculated analytically in the compression zone.
		The burning-rate eigenvalue problem is also considered to derive a formula of laminar flame speed, which is coupled with differential equation for temperature in the reaction zone.
		Then, the dependence of laminar flame speed on thermal-diffusive parameters, accompanied with hydrodynamic properties brought by the compression effect, becomes clear.
		The non-zero gradient condition of temperature on the burned side of a flame front is obtained for non-zero values of Mach number.
		Opposed to the monotonic profile of temperature in the zero-Mach-number limit, the condition of negative temperature gradient brings the non-monotonic behavior of temperature.
		It is found that the smallest value of laminar flame speed is achieved for some negative small value of temperature gradient.
	\end{abstract}
	

\end{center}


\section{INTRODUCTION}
The adiabatic profile of a premixed flame front has been studied usually based on the assumption of zero temperature gradient at the burned-side edge of the front \cite{Sivashinsky76}.
In the zero-Mach-number limit, this assumption is consistent with the results obtained in experiments \cite{Lewis87}.
In this case, the large activation energy asymptotics is a powerful method to understand the flow properties inside and outside the reaction zone \cite{Zeldovich38a,Zeldovich38b,Sivashinsky75,Sivashinsky77,Matkowsky79}.
The boundary conditions across the reaction zone have been derived based on the method of matched asymptotic expansions \cite{Fife88}.
These conditions are in turn used for the study of preheat zone inside a flame front \cite{Clavin82,Pelce82,Matalon82,Matalon03}, which gives the correction of Darrieus-Landau instability (DLI) \cite{Darrieus38,Landau44} due to the effects of curvature of a flame front \cite{Markstein51} or tangential velocity \cite{Eckhaus59,Eckhaus61}.

In the study of reaction zone, the exothermic reaction term is usually assumed to be the Arrhenius type, which has an explicit dependence on temperature \cite{Spalding57b,Spalding57c}.
In this case, it is possible to derive the parameter dependence of laminar flame speed by considering the burning-rate eigenvalue problem \cite{Friedman53,Karman57,Bush70}.
Due to the low-Mach-number limit, the thermal-diffusive properties are decoupled completely from the hydrodynamic flow inside the reaction zone \cite{Joulin98}.
Therefore, only the heat-conduction and mass-fraction equations are calculated to give the asymptotic solution of burning-rate eigenvalue under the assumption of zero temperature gradient on the burned side.
On the other hand, the non-zero gradient of temperature has also been studied \cite{Linan74,Law06}.
In this study, we consider the relation between the temperature gradient and laminar flame speed.
This relation may be important in some situations related to the deflagration-to-detonation transition (DDT), where the non-zero values of temperature gradient are observed for a fast propagating deflagration or a reaction wave behind the shock of detonation \cite{Clavin16}.

The DDT is a serious problem in some scientific fields: for example, the terrestrial combustion in narrow channels \cite{Kurdyumov16} and mines \cite{Demir18}, or the thermonuclear explosions which are related to the understanding of local expansion rate of the universe \cite{Hillebrandt00}.
The mechanism of DDT remains unclear and the numerical analysis of flow properties has been performed in channels \cite{Kagan15,Kurdyumov17}.
In confined circumstances, the acceleration of flame propagation speed is induced by the effects of thermal expansion and wall frictions \cite{Brailovsky00,Demirgok15}.
The increase of area of flame surface is essential for the rise of burning rate \cite{Karlovitz53,Buckmaster79,Matalon83}.
This is mainly due to the corrugation of front invoked by the DLI \cite{Bychkov08,Modestov09}.
The theoretical study about the effect of gas compression on the DLI in tubes has been performed by small Mach number expansions \cite{Bychkov10,Valiev13}.

The DDT is also important when we consider the Type Ia supernova (SN Ia) in the process of thermonuclear explosions \cite{Wheeler90,Ropke18}.
The SN Ia has attracted many researchers' interest to its theoretical analysis.
This is because the SN Ia may be modelled by the classical thermal theory of flames \cite{Mallard83}: for example, the burning of premixed carbon-oxygen fuel \cite{Hoyle60,Boisseau96}.
At the early stage of explosions, the propagation of flames is observed with the Mach number in the range of $0.01$-$0.1$ \cite{Woosley01,Almgren06,Pfannes10}.
In this case, the spatial distribution of temperature has been studied numerically based on the deflagration model \cite{Glazyrin13}, detonation model \cite{Khokhlov89,Sharpe99} and reconstruction model \cite{Townsley16}.
In every work, the non-monotonic profile of temperature is reported with its maximum value is achieved inside the reaction zone.
As for the acceleration of thermonuclear flames, the Rayleigh-Taylor instability due to buoyancy effect is an essential factor inducing the corrugated fronts \cite{Bychkov07}.
Therefore, the multidimensional structure, which lead to the formation of turbulence, must be considered in the process of DDT.
Although the curvature effect is neglected in this study, our results may be adopted to the local estimation of properties of flames propagating with small Mach numbers at the early stage of DDT.
Indeed, our analysis is performed in the microscopic scale expressed by the order of thickness of reaction zone.

In the previous work \cite{Wada21apr}, the compression zone, whose thickness is $O({M\hspace{-1.5pt}a}^2)$, was introduced behind a flame front under ${M\hspace{-1.5pt}a}^2\ll1$ with ${M\hspace{-1.5pt}a}$ denoting the Mach number, which is defined by the ratio of laminar flame speed to local adiabatic sound speed: similar approaches are found, for example, in \cite{Kapila83,Bush99}.
This approach enables us to capture the non-adiabatic temperature profile with the temperature gradient at the burned-side edge being no longer zero.
Especially, such a non-adiabatic profile is described by inner solutions inside the compression zone, which are expressed by use of Lambert $W$ function \cite{Corless96}.
However, the strong ignition temperature approximation was employed in order to obtain analytical solutions inside a flame front, which consists of the preheat and reaction zones \cite{Curtiss59,Hirschfelder61a,Hirschfelder61b,Bowen67}.
Therefore, the temperature dependence of reaction term was ignored and the cold boundary difficulty was unresolved.
Instead, in this work, the reaction term is expressed by the Arrhenius type and its temperature dependence is accounted for \cite{Glassman16,Wada21oct}.
By employing the large activation energy asymptotics, the cold boundary difficulty is resolved \cite{Williams85}.

In this study, the order of Mach number is assumed to be squared root of the inverse of non-dimensional activation energy.
Thanks to this specific order, under the large values of activation energy, the effect of gas compression is able to be captured in the reaction zone as well as in the compression zone.
As a result, the condition of temperature gradient at the burned-side edge of a flame front is simply described by the difference between the flame temperature and its adiabatic value.
Besides, the burning-rate eigenvalue contains the influence of not only thermal-diffusive properties but also hydrodynamic flow.
In this case, the equations of heat conduction and burning-rate eigenvalue need to be computed simultaneously to obtain the solutions of temperature and laminar flame speed.

The non-zero temperature gradient for small Mach numbers brings two specific features of spatial temperature distribution.
For the positive temperature gradient, the monotonic increase of temperature is observed, which is similar with that in the adiabatic case.
In this case, the augmentation of viscous effect, which is denoted by the Prandtl number $Pr$, induces the drop of temperature.
On the other hand, in the case of negative temperature gradient, the non-monotonic temperature profile is captured with its maximum value taken inside the reaction zone.
Then, the increase of $Pr$ causes that of temperature.

\section{PRELIMINARIES}\label{sec:Preliminaries}
\subsection{Governing Equations}\label{subsec:Governing equations}
We consider a planar premixed flame front freely propagating in the negative z-direction.
The coordinate $z$ is non-dimensionalised by the thermal thickness of a flame front, or the length scale of preheat zone \cite{Sivashinsky76}.
The density $R$, velocity of gases $V$, pressure $P$, temperature $T$ and mass fraction of deficient species $Y$ are made dimensionless by use of their values at a position far from a flame front, $z\to-\infty$, denoted by $\tilde{R}_{-\infty}$, $\tilde{S}_{L}$, $\tilde{P}_{-\infty}$, $\tilde{T}_{-\infty}$ and $\tilde{Y}_{-\infty}$.
Especially, $\tilde{S}_{L}$ is called the laminar flame speed, or the burning velocity, which is the propagation velocity of a planar flame front into a quiescent fuel mixture \cite{Williams85}.
Hereafter, the variables with tilde symbol stand for dimensional ones.
\begin{alignat}{5}
	R &= \frac{\tilde{R}}{\tilde{R}_{-\infty}},& V &= \frac{\tilde{V}}{\tilde{S}_{L}},& P &= \frac{\tilde{P}}{\tilde{P}_{-\infty}},& T &= \frac{\tilde{T}}{\tilde{T}_{-\infty}},& Y &= \frac{\tilde{Y}}{\tilde{Y}_{-\infty}},
	\notag \\
	\tilde{l}_d &= \frac{\tilde{D}_{th}}{\tilde{S}_{L}},\quad& Le &= \frac{\tilde{D}_{th}}{\tilde{D}},\quad& Pr &=\frac{\tilde{\mu}\tilde{c}_p}{\tilde{\lambda}},\quad& \tilde{D}_{th} &= \frac{\tilde{\lambda}}{\tilde{R}_{-\infty}\tilde{c}_p}, \quad&M\hspace{-1.5pt}a &= \frac{\tilde{S}_{L}}{\tilde{c}_s}, \ \
	\label{nondimensionalization}
\end{alignat}
where $\tilde{l}_d$, $\tilde{D}_{th}$, $\tilde{D}$, $\tilde{\mu}$, $\tilde{c}_p$ and $\tilde{\lambda}$ are the length scale of flame front, the thermal diffusivity, the diffusion coefficient of a deficient species, the coefficient of viscosity, the specific heat at constant pressure of the mixture and the thermal conductivity. 
We assume that $\tilde{D}$, $\tilde{\mu}$, $\tilde{c}_p$ and $\tilde{\lambda}$ are all constant.
The adiabatic sound speed is defined by $\tilde{c}_s=(\gamma \tilde{P}_{-\infty}/\tilde{R}_{-\infty})^{1/2}$.
The specific heat ratio, Mach number, Prandtl number and Lewis number are denoted by $\gamma$, $M\hspace{-1.5pt}a$, $Pr$ and $Le$, respectively.

Because of (\ref{nondimensionalization}), we have the following far-field boundary conditions.
\begin{align}
	R=V=P=T=Y=1 \quad \mbox{as}\quad z\to-\infty.\label{bdc:far_field}
\end{align}
For a steady one-dimensional flow, the conservation law of mass tells us that the mass flux $M$ is constant in the entire region including the inside of flame front.
Then, the boundary condition (\ref{bdc:far_field}) leads to
\begin{align}
	M = RV = 1.\label{mass_flux_unity}
\end{align}
Accounting for (\ref{mass_flux_unity}), we have the following non-dimensional governing equations.
\begin{align}
	\frac{d V}{d z} &= -\frac{1}{\gamma {M\hspace{-1.5pt}a}^2}\frac{d P}{d z} + \frac{4}{3} Pr\frac{d^2 V}{d z^2},\label{eqn:momentum}\\
	\frac{d T}{d z} &= \frac{d^2 T}{d z^2} + \frac{\gamma - 1}{\gamma}V\frac{d P}{d z}+\frac{4}{3}(\gamma-1)Pr{M\hspace{-1.5pt}a}^2\left(\frac{d V}{d z}\right)^2 + qQ,\label{eqn:heat}\\
	\frac{d Y}{d z} &= \frac{1}{Le}\frac{d^2 Y}{d z^2} - Q ,\label{eqn:fuel}\\
	P &= RT ,\label{eqn:state}
\end{align}
where $q=\tilde{q}\tilde{Y}_{-\infty}/(\tilde{c}_p\tilde{T}_{-\infty})$ and $Q$ are the non-dimensional heat release and reaction rate with $\tilde{q}$ being the dimensional heat release.
We assume that the deficient species is depleted inside a flame front and does not exist on the burned side.
\begin{align}
	Y \equiv 0\quad (z\geq0).\label{assm:Y=0}
\end{align}
We note that the assumption (\ref{assm:Y=0}) contains the equality in the range of $z$.
This means that the burned-side edge of a flame front is located at $z=0$.

The reaction rate $Q$ is assumed to be the Arrhenius type:
\begin{equation}
	Q = RY\frac{\tilde{A} \tilde{D}_{th}}{\tilde{S}_{L}^2}\exp\left(-\frac{N}{T}\right),
	\label{eqn:def-Q}
\end{equation}
where $\tilde{A}$ is the frequency factor and assumed to be constant. The non-dimensional activation energy is expressed by $N = \tilde{E} / \tilde{R}_g \tilde{T}_{-\infty}$ with the dimensional activation energy $\tilde{E}$ and the universal gas constant $\tilde{R}_g$.
In order to avoid the cold boundary difficulty in the limit of large activation energy $N\gg1$, $Q$ is transformed into the following form \cite{Matalon82,Williams85}.
\begin{eqnarray}
	&Q& = \frac{\Lambda R Y}{\delta}\left(\frac{N}{T_b^2}\right)^2\exp\left(N\frac{T-T_b}{T_bT}\right),
	\label{Q}
\end{eqnarray}
where
\begin{eqnarray}
	\Lambda = \frac{ \tilde{A} \tilde{D}_{th}}{\tilde{S}_{L}^2}\left(\frac{T_b^2}{N}\right)^2\exp\left(-\frac{N}{T_b}\right).
	\label{Lambda}
\end{eqnarray}
The coefficient $\Lambda$ plays a role of the eigenvalue which determines the laminar flame speed $\tilde{S}_L$ and is called the burning-rate eigenvalue \cite{Zeldovich38a,Zeldovich38b,Friedman53,Karman57,Bush70,Williams85}.
Under the large-activation-energy asymptotics, $N\gg1$, we introduce a small parameter $\epsilon$, which characterizes the thickness of reaction zone.
\begin{eqnarray}
	\epsilon=T_b^2/N= (T_b-1)/\beta,\label{def:epsilon}
\end{eqnarray}
where the Zel'dovich number $\beta$ \cite{Glassman16} is defined by
\begin{eqnarray}
	\beta = \frac{\tilde{E}}{\tilde{R}_g\tilde{T}_b}\frac{\tilde{T}_b-\tilde{T}_{-\infty}}{\tilde{T}_b}.\label{def:beta}
\end{eqnarray}
In comparing our result with that of previous works, it may be helpful to rewrite $\epsilon$ into the form expressed by the Zel'dovich number.
The flame temperature $T_b$ is evaluated at the burned-side edge of a flame front under $\epsilon\ll1$, and is different from the adiabatic flame temperature $T_{ad}$. 
\begin{remark}
	The essential assumption, which determines the adiabatic profile of a flame front, is that the temperature gradient is zero at $z=0$.
	Then, the adiabatic flame temperature is given by $T_{ad}=1+q$ \cite{Sivashinsky77,Matalon82}.
\end{remark}
In this study, the above assumption of zero-gradient of temperature at $z=0$ is removed.
As a result, the non-monotonic behavior of temperature is captured for its negative gradients with the maximum value of temperature taken inside the reaction zone.
Such a temperature profile may be observed in the context of thermonuclear explosions \cite{Glazyrin13,Khokhlov89,Sharpe99,Townsley16}.

By use of (\ref{def:epsilon}), the burning-rate eigenvalue (\ref{Lambda}) is rewritten as
\begin{align}
	\Lambda = \frac{\tilde{A} \tilde{D}_{th}}{\tilde{S}_{L}^2}\epsilon^2\exp\left(-\frac{T_b}{\epsilon}\right).\label{Lambda_rewritten}
\end{align}
We seek for the burning-rate eigenvalue in the form of power series with respect to $\epsilon\;(\ll1)$.
\begin{equation}
	\Lambda = \Lambda_{0} + O(\epsilon).\label{Lambda_expansion_epsilon}
\end{equation}

All quantities are also supposed to be expanded asymptotically with respect to $\epsilon$.
In the outer regions ($z<0,z>0$),
\begin{subequations}
	\begin{alignat}{3}
		R &= {R}_{0} + \epsilon{R}_{1},\quad&V &= {V}_{0} + \epsilon{V}_{1},\quad& P &= {P}_{0} + \epsilon{P}_{1},\label{epsilon_expansions_outer_RWP} \\
		T &= {T}_{0} + \epsilon{T}_{1},& Y &= {Y}_{0} + \epsilon{Y}_{1},&&\label{epsilon_expansions_outer_TY} 
	\end{alignat}\label{epsilon_expansions_outer}
\end{subequations}
in the reaction zone ($-\infty<\eta<0$),
\begin{subequations}
	\begin{alignat}{3}
		R &= \hat{\rho}_{0} + \epsilon\hat{\rho}_{1},&V &= \hat{v}_{0} + \epsilon\hat{v}_{1},\;& P &= \hat{p}_{0} + \epsilon\hat{p}_{1},\label{epsilon_expansions_reaction_RWP}\\
		T &= \hat{\theta}_{0} + \epsilon\hat{\theta}_{1} + \epsilon^2\hat{\theta}_{2},\;& Y &= \hat{c}_{0} + \epsilon\hat{c}_{1} + \epsilon^2\hat{c}_{2},&&\label{epsilon_expansions_reaction_TY}
	\end{alignat}\label{epsilon_expansions_reaction}
\end{subequations}
and in the compression zone ($0<\xi<+\infty$),
\begin{subequations}
	\begin{alignat}{3}
		R &= \check{\rho}_{0} + \epsilon\check{\rho}_{1},&V &= \check{v}_{0} + \epsilon\check{v}_{1},\;& P &= \check{p}_{0} + \epsilon\check{p}_{1},\label{epsilon_expansions_compression_RWP}\\
		T &= \check{\theta}_{0} + \epsilon\check{\theta}_{1} + \epsilon^2\check{\theta}_{2},\;& Y &= \check{c}_{0} + \epsilon\check{c}_{1} + \epsilon^2\check{c}_{2}.&&\label{epsilon_expansions_compression_TY}
	\end{alignat}\label{epsilon_expansions_compression}
\end{subequations}

\subsection{The Order of Mach Number}
In this study, we consider the following order of Mach number with positive constant $b=O(1)$.
\begin{align}
	{M\hspace{-1.5pt}a} = b\sqrt{\epsilon}.\label{def:b}
\end{align}
The assumption (\ref{def:b}) enables us to comprehend analytically the effect of compressibility, or finite Mach number, on the temperature profile of a flame front.
Indeed, the pressure and dissipation terms in the heat-conduction equation emerge as a consequence of non-zero values of Mach number, or $b>0$.
Then, the adiabatic profile of a flame front is greatly changed: for example, the non-monotonic behavior of temperature is observed.
Especially, the variation of temperature in the compression zone is described by the inner asymptotic solution expressed by Lambert $W$ function.
The similar result is obtained in \cite{Wada21apr} based on the strong ignition temperature approximation, which ignored the temperature dependence of reaction rate.
Contrary to the previous work, we systematically deal with the reaction rate represented by the Arrhenius type \cite{Spalding57b,Spalding57c} by use of the large activation energy asymptotics.
As a result, the burning-rate eigenvalue is found to be affected by the compressibility.
In other words, the laminar flame speed reflects the non-adiabatic feature of a flame front.

\subsection{Notation in Reaction and Compression Zones}
\label{subsec:Notation in reaction and compression zones}
We introduce the reaction zone inside a flame front and the compression zone behind it based on the large activation energy asymptotics under $\epsilon\ll1$.
The inner variables, or the stretched variables, are introduced as
\begin{align}
	z = \epsilon\eta\quad(-\infty<\eta<0),\quad z = {M\hspace{-1.5pt}a}^2\xi = b^2\epsilon\xi\quad(0<\xi<+\infty),\label{def:inner_variables}
\end{align}
where $\eta$ and $\xi$ represent the coordinates in the reaction and compression zones respectively, and $z$ is the variable in outer regions, which consist of the unburned ($z<0$) and burned ($z>0$) sides.
The study of inner regions is made to calculate inner solutions, which are given by (\ref{epsilon_expansions_reaction}) and (\ref{epsilon_expansions_compression}), and the boundary conditions across them, which are in turn used to seek for outer solutions.

\subsubsection{Matching Conditions Between Outer and Inner Solutions}\label{subsubsec:Matching conditions between outer and inner solutions}
For convenience in the later calculation, we recall the matching conditions between the inner and outer solutions in the overlapping regions, where $\eta\to-\infty$ and $z\to0-$ or $\xi\to+\infty$ and $z\to0+$ as $\epsilon\to0$.
Let $\Phi$ be any function and its asymptotic expansions with respect to $\epsilon$ are assumed to be
\begin{alignat*}{2}
	\Phi &= {\Phi}_{0}(z) + \epsilon{\Phi}_{1}(z) + \epsilon^2{\Phi}_{2}(z) + O(\epsilon^3)\quad& &(z<0,z>0),\\
	\Phi &= \hat{\phi}_{0}(\eta) + \epsilon\hat{\phi}_{1}(\eta) + \epsilon^2\hat{\phi}_{2}(\eta) + O(\epsilon^3)\quad& &(-\infty<\eta<0),\\
	\Phi &= \check{\phi}_{0}(\xi) + \epsilon\check{\phi}_{1}(\xi) + \epsilon^2\check{\phi}_{2}(\xi) + O(\epsilon^3)\quad& &(0<\xi<+\infty).
\end{alignat*}
We assume that the inner and outer solutions and their derivatives are continuous in the overlapping regions.
Then, according to \cite{Fife88}, the matching conditions are given by, between the reaction and unburned zones,
\begin{subequations}
	\begin{align}
		\hat{\phi}_{0}|_{\eta\to-\infty} &= {\Phi}_{0}|_{z\to 0-},\quad
		\hat{\phi}_{1}|_{\eta\to-\infty} = {\Phi}_{1}|_{z\to 0-} + \eta\frac{d {\Phi}_{0}}{d z}\Big|_{z\to 0-},\label{matching_conditions_reaction}\\
		\frac{d \hat{\phi}_{0}}{d \eta}\Big|_{\eta\to-\infty} &= 0,\quad\quad\quad\;
		\frac{d \hat{\phi}_{1}}{d \eta}\Big|_{\eta\to-\infty} = \frac{d {\Phi}_{0}}{d z}\Big|_{z\to 0-},\label{matching_conditions_d_reaction}\\
		\frac{d \hat{\phi}_{2}}{d \eta}\Big|_{\eta\to-\infty} &= \frac{d {\Phi}_{1}}{d z}\Big|_{z\to 0-} + \eta\frac{d^2 {\Phi}_{0}}{d z^2}\Big|_{z\to 0-},\label{matching_conditions_d2_reaction}\\
		\frac{d^2 \hat{\phi}_{1}}{d \eta^2}\Big|_{\eta\to-\infty} &= 0,\quad\quad\quad\;
		\frac{d^2 \hat{\phi}_{2}}{d \eta^2}\Big|_{\eta\to-\infty} = \frac{d^2 {\Phi}_{0}}{d z^2}\Big|_{z\to 0-}\label{matching_conditions_d_reaction_2}
	\end{align}\label{matching_conditions_all_reaction}
\end{subequations}
and between the compression and burned zones,
\begin{subequations}
	\begin{align}
		\check{\phi}_{0}|_{\xi\to+\infty} &= {\Phi}_{0}|_{z\to 0+},\quad
		\check{\phi}_{1}|_{\xi\to+\infty} = {\Phi}_{1}|_{z\to 0+} + b^2\xi\frac{d {\Phi}_{0}}{d z}\Big|_{z\to 0+},\label{matching_conditions_compression}\\
		\frac{d \check{\phi}_{0}}{d \xi}\Big|_{\xi\to+\infty} &= 0,\quad\quad\quad\;
		\frac{d \check{\phi}_{1}}{d \xi}\Big|_{\xi\to+\infty} = b^2\frac{d {\Phi}_{0}}{d z}\Big|_{z\to 0+},\label{matching_conditions_d_compression}\\
		\frac{d \check{\phi}_{2}}{d \xi}\Big|_{\xi\to+\infty} &= b^2\frac{d {\Phi}_{1}}{d z}\Big|_{z\to 0+} + b^4\xi\frac{d^2 {\Phi}_{0}}{d z^2}\Big|_{z\to 0+},\label{matching_conditions_d2_compression}\\
		\frac{d^2 \check{\phi}_{2}}{d \xi^2}\Big|_{\xi\to+\infty} &= b^4\frac{d^2 {\Phi}_{0}}{d z^2}\Big|_{z\to 0+}.\label{matching_conditions_d3_compression}
	\end{align}\label{matching_conditions_all_compression}
\end{subequations}
In addition, we impose the following assumptions.
\begin{remark}
	We assume that all quantities are continuous at the burned-side edge of a flame front, $z=0$.
	\begin{align}
		\hat{\phi}_{i}|_{\eta\to0-} = \check{\phi}_{i}|_{\xi\to0+}\quad(i=0,1).\label{continuity_conditions}
	\end{align}\label{rem:continuity_eta=0}
\end{remark}
\begin{remark}
	We further assume that the derivatives of inner solutions at the order higher than $\epsilon$ are continuous at $z=0$.
	This condition is not imposed on leading-order solutions, because we assume inner solutions of temperature and mass fraction is constant inside of reaction zone, and in this case, only constant solutions are admissible in reaction zone.
	\begin{align}
		\frac{d \hat{\phi}_i}{d \eta}\Big|_{\eta\to0-} = \frac{1}{b^2}\frac{d \check{\phi}_i}{d \xi}\Big|_{\xi\to0+},\quad(i=1,2).\label{assm:heat_fuel_derivative}
	\end{align}\label{rem:derivative_continuity}
\end{remark}
For simplicity, we will use the following notation.
\begin{align*}
	{\Phi}_{0}|_{z\to 0\pm} = {\Phi}_{0\pm},\quad \frac{d {\Phi}_{0}}{d z}\Big|_{z\to 0\pm} = \frac{d {\Phi}_{0}}{d z}\Big|_{\pm}.
\end{align*}
The same things are true for the higher order terms.

\section{INNER SOLUTIONS}
\label{sec:Inner solutions}
At first, we study the inner structure of compression zone by employing the assumption of Mach number (\ref{def:b}) and the matching conditions (\ref{matching_conditions_all_compression}).
After that, the inner solutions of reaction zone are calculated with the help of matching conditions (\ref{matching_conditions_all_reaction}).
Then, the solutions in reaction and compression zones are connected by use of continuity conditions (\ref{continuity_conditions}) and (\ref{assm:heat_fuel_derivative}) to derive the boundary conditions for outer solutions.

According to previous works \cite{Karman57,Bush70,Matkowsky79}, we further impose the following assumption.
\begin{remark}
	We assume the following solutions inside reaction zone ($-\infty<\eta<0$).
	\begin{align}
		\hat{\theta}_0 = T_b,\quad \hat{c}_0 = 0.\label{eqn:heat_fuel_reaction_zone_0}
	\end{align}
	This means that the temperature is sufficiently high in the entire region of reaction zone and nearly equal to its value at the burned-side edge of a flame front.
	Correspondingly, at the leading order of $\epsilon$, the mass fraction also has its value on the burned side.
	\label{rem:heat_fuel_assumptions}
\end{remark}

\subsection{Compression Zone $(0<\xi<+\infty)$}
\label{subsec:Compression zone}
\subsubsection{$O(\epsilon^0)$ Solutions in Compression Zone}
\label{subsubsec:O(epsilon0) solutions in compression zone}
From the assumption (\ref{assm:Y=0}), we have
\begin{align}
	\check{c}_i = 0\quad (i=0,1,2\ldots).\label{sol:c_compression}
\end{align}
Collecting $O(\epsilon^0)$ terms, governing equations (\ref{mass_flux_unity})-(\ref{eqn:heat}) and (\ref{eqn:state}) become
\begin{align}
	\check{\rho}_0\check{v}_0 &= 1,\label{eqn:mass_compression_zone_0}\\
	0 &= -\frac{1}{\gamma}\frac{d \check{p}_0}{d \xi} + \frac{4}{3} Pr\frac{d^2 \check{v}_0}{d \xi^2},\label{eqn:momentum_compression_zone_0}\\
	0 &= \frac{d^2 \check{\theta}_0}{d \xi^2},\label{eqn:heat_compression_zone_0}\\
	\check{p}_0 &= \check{\rho}_0\check{\theta}_0.\label{eqn:state_compression_zone_0}
\end{align}
From (\ref{eqn:heat_compression_zone_0}), accounting for (\ref{matching_conditions_d_reaction}), (\ref{continuity_conditions}) and (\ref{eqn:heat_fuel_reaction_zone_0}), we readily find that
\begin{align}
	\check{\theta}_0 = T_b.\label{t_0_compression}
\end{align}
Integrating (\ref{eqn:momentum_compression_zone_0}), with the help of (\ref{matching_conditions_all_compression}), we get
\begin{align}
	\check{p}_{0} = \nu\frac{d \check{v}_{0}}{d \xi} + P_{0+},\label{pressure_0_integral}
\end{align}
where we put
\begin{align}
	\nu = \frac{4}{3}\gamma Pr.\label{def:nu}
\end{align}
We find that the pressure is constant for inviscid flow, $Pr=0$.
Then, the velocity also becomes constant and only the adiabatic profile of a premixed flame front is observed.

From (\ref{matching_conditions_compression}), (\ref{eqn:mass_compression_zone_0}), (\ref{eqn:state_compression_zone_0}) and (\ref{t_0_compression}), in the limit of $\xi\to+\infty$, we find
\begin{align}
	\check{\theta}_{0} = T_b = {P}_{0+}{V}_{0+}.\label{tc_inner_0}
\end{align}
Substituting (\ref{eqn:state_compression_zone_0}) into (\ref{pressure_0_integral}), with the help of (\ref{tc_inner_0}), we obtain the following differential equation for $\check{v}_{0}$ ($>0$).
\begin{align}
	\frac{\check{v}_{0}}{\check{v}_{0}-{V}_{0+}}\frac{d \check{v}_{0}}{d \xi} = -\frac{{P}_{0+}}{\nu}.\label{eqn:nonlinear_v0}
\end{align}
By use of the upper branch of Lambert $W$ function \cite{Curtiss59}, which is expressed by $y = W_0\left(x\right)$ as the inverse relation of $y=x\mathrm{e}^x$ for real numbers $x$ and $y$ ($x\geq-1/e, y\geq-1$), we get the solution of (\ref{eqn:nonlinear_v0}) as follows \cite{Wada21apr}.
\begin{align}
	\check{v}_{0} = {V}_{0+}\left(1+W\right),\label{sol:v0_compression}
\end{align}
where
\begin{align}
	W &= W_0\left(\alpha\exp\left(-\frac{1}{\nu}\frac{{P}_{0+}}{{V}_{0+}}\xi+\alpha\right)\right),\label{W=W0}\\
	\alpha &= W|_{\xi=0+} = \frac{\check{v}_{0+}}{{V}_{0+}}-1.\label{def:alpha}
\end{align}
Substituting (\ref{sol:v0_compression}) into (\ref{pressure_0_integral}), we obtain
\begin{align}
	\check{p}_{0} = {P}_{0+}(1+W)^{-1}.\label{sol:p0_compression}
\end{align}

\subsubsection{$O(\epsilon)$ Solutions in Compression Zone}
\label{subsubsec:epsilon solution of temperature in compression zone}
At $O(\epsilon)$, the heat-conduction equation (\ref{eqn:heat}) reads
\begin{align}
	0 &= \frac{d^2 \check{\theta}_1}{d \xi^2} + b^2\frac{\gamma - 1}{\gamma}\left(\check{v}_0\frac{d \check{p}_0}{d \xi} + \nu\left(\frac{d \check{v}_{0}}{d \xi}\right)^2\right).\label{eqn:heat_compression_zone_1}
\end{align}
Accounting for (\ref{matching_conditions_d_compression}), (\ref{sol:v0_compression}) and (\ref{sol:p0_compression}), the integration of (\ref{eqn:heat_compression_zone_1}) leads to
\begin{align}
	\frac{d \check{\theta}_{1}}{d \xi} = b^2\frac{\gamma-1}{\gamma}\check{\theta}_{0}W+b^2\frac{d {T}_{0}}{d z}\Big|_{+}.\label{eqn:heat_reaction_1_positive_integrate}
\end{align}
Integrating (\ref{eqn:heat_reaction_1_positive_integrate}), by use of (\ref{matching_conditions_compression}), we get
\begin{align}
	\check{\theta}_{1} = {T}_{1+} + \xi b^2\frac{d {T}_{0}}{d z}\Big|_{+}-b^2\frac{\gamma-1}{\gamma}\nu{V}_{0+}^2W\left(1+\frac{W}{2}\right).\label{sol:theta1_compression}
\end{align}

Next, we calculate $O(\epsilon)$ solutions of velocity and pressure in the compression zone.
Collecting $O(\epsilon)$ terms in governing equations (\ref{mass_flux_unity}), (\ref{eqn:momentum}) and (\ref{eqn:state}), we have
\begin{align}
	0 &= \check{\rho}_{1}\check{v}_{0}+\check{\rho}_{0}\check{v}_{1},\label{eqn:mass_compression_zone_1}\\
	\gamma b^2\frac{d \check{v}_0}{d \xi} &= -\frac{d \check{p}_1}{d \xi} + \nu\frac{d^2 \check{v}_1}{d \xi^2},\label{eqn:momentum_compression_zone_1}\\
	\check{p}_1 &= \check{\rho}_1\check{\theta}_0 + \check{\rho}_0\check{\theta}_1.\label{eqn:state_compression_zone_1}
\end{align}
From (\ref{eqn:mass_compression_zone_0}), (\ref{eqn:state_compression_zone_0}), (\ref{eqn:mass_compression_zone_1}) and (\ref{eqn:state_compression_zone_1}), $\check{p}_1$ is calculated by
\begin{align}
	\check{p}_1 = -\frac{\check{p}_0}{\check{v}_0}\check{v}_1 + \frac{\check{\theta}_1}{\check{v}_0}.\label{sol:p1_compression_zone}
\end{align}
Substituting (\ref{sol:p1_compression_zone}) into the integration of (\ref{eqn:momentum_compression_zone_1}), by use of the matching conditions (\ref{matching_conditions_compression}) and (\ref{matching_conditions_d_compression}) with the matching relation $dP_0/dz|_+=0$ which is found by taking $\xi\to+\infty$ in (\ref{eqn:momentum_compression_zone_1}), we obtain the following differential equation for $\check{v}_1$.
\begin{align}
	\nu\frac{d \check{v}_1}{d \xi} + \frac{\check{p}_0}{\check{v}_0}\check{v}_1 = \frac{\check{\theta}_1}{\check{v}_0} + \gamma b^2\check{v}_0 + b^2\nu\frac{d V_0}{d z}\Big|_+ - P_{1+} - \gamma b^2V_{0+},\label{eqn:v1_compression_zone_1}
\end{align}
Solving (\ref{eqn:v1_compression_zone_1}), by use of solutions (\ref{sol:v0_compression}), (\ref{sol:p0_compression}) and (\ref{sol:theta1_compression}), we get
\begin{align}
	\check{v}_{1} &= \frac{1}{{P}_{0+}}\frac{W}{1+W}\left\{\check{v}_{1c}+({T}_{1+}-{P}_{1+}{V}_{0+})\left(\frac{1}{W}-\ln{|W|}\right)\notag\right.\\
	&\left.\quad-b^2{V}_{0+}^2\frac{1+\gamma-B}{4}W(2+W)+\frac{{P}_{0+}}{\nu}(b^2{V}_{0+}(1-B)-{P}_{1+})\xi\notag\right.\\
	&\left.\quad+b^2\frac{d T_0}{d z}\Big|_+\left(\left(\frac{1}{W}-1-\ln{|W|}\right)\xi-\nu\frac{V_{0+}}{P_{0+}}\left(W+\frac{\ln{|W|}}{2}\right)\ln{|W|}\right)\notag\right.\\
	&\left.\quad-b^2V_{0+}\nu\frac{d V_0}{d z}\Big|_+(W+2\ln{|W|})
	\right\},\label{sol:v1_compression_zone}
\end{align}
where
\begin{align}
	B &= \frac{\gamma-1}{\gamma}(\nu-\gamma),\label{def:B}\\
	\check{v}_{1c} &= \frac{1+\alpha}{\alpha}{P}_{0+}\check{v}_{1+}-({T}_{1+}-{P}_{1+}{V}_{0+})\left(\frac{1}{\alpha}-\ln{|\alpha|}\right)\notag\\
	&\quad+b^2{V}_{0+}^2\frac{1+\gamma-B}{4}\alpha(2+\alpha)+b^2\frac{V_{0+}}{P_{0+}}\nu\frac{d T_0}{d z}\Big|_+\left(\alpha+\frac{\ln{|\alpha|}}{2}\right)\ln{|\alpha|}\notag\\
	&\quad+b^2V_{0+}\nu\frac{d V_0}{d z}\Big|_+(\alpha+2\ln{|\alpha|}).\label{v1c_compression_zone}
\end{align}
Solutions (\ref{sol:p1_compression_zone}) and (\ref{sol:v1_compression_zone}) are used to calculate the heat-conduction equation at $O(\epsilon^2)$ in Section \ref{subsubsec:epsilon2 solutions in compression zone}.
Besides, they are matched with inner solutions of reaction zone to determine the values of $T_{1+}$ and $P_{1+}$.

\subsubsection{Heat-Conduction Equation at $O(\epsilon^2)$ in Compression Zone}
\label{subsubsec:epsilon2 solutions in compression zone}
We consider the heat-conduction equation (\ref{eqn:heat}) at $O(\epsilon^2)$ as follows.
\begin{align}
	b^2\frac{d \check{\theta}_1}{d \xi} = \frac{d^2 \check{\theta}_2}{d \xi^2} + b^2\frac{\gamma - 1}{\gamma}\left(\check{v}_1\frac{d \check{p}_0}{d \xi}+\check{v}_0\frac{d \check{p}_1}{d \xi}+2\nu\frac{d \check{v}_{0}}{d \xi}\frac{d \check{v}_{1}}{d \xi}\right).\label{eqn:heat_fuel_compression_zone_2}
\end{align}
The integration of (\ref{eqn:heat_fuel_compression_zone_2}), by use of solutions (\ref{sol:v0_compression}), (\ref{sol:p0_compression}), (\ref{sol:p1_compression_zone}) and (\ref{sol:v1_compression_zone}), leads to
\begin{align}
	&\frac{\gamma}{\gamma-1}\check{\theta}_1 - \frac{1}{b^2}\frac{\gamma}{\gamma-1}\frac{d \check{\theta}_2}{d \xi} + \check{\theta}_{1c}\notag\\
	&= \frac{\check{v}_{1c}}{1+W} + T_{1+}\frac{W}{1+W}\left(1+\ln{|W|}\right)
	-P_{1+}V_{0+}\left\{\frac{W\ln{|W|}-1}{1+W}+2(1+W)\notag\right.\\
	&\left.\quad+\ln{|W|}+\frac{P_{0+}}{\nu V_{0+}}\frac{\xi}{1+W}\right\}
	+ b^2V_{0+}^2(1-B)\left\{\frac{1}{4(1+W)}+\frac{3}{4}W(3+W)\notag\right.\\
	&\left.\quad+\ln{|W|}+\frac{P_{0+}}{\nu V_{0+}}\frac{\xi}{1+W}\right\}
	+ b^2V_{0+}^2\frac{\gamma}{4}\left(\frac{1}{1+W}+(W+1)(W-4)\right)\notag\\
	&\quad+ b^2\frac{V_{0+}}{P_{0+}}\nu\frac{d T_0}{d z}\Big|_{+}\left\{\frac{1}{1+W}-W+\frac{W\ln{|W|}}{1+W}\left(W+\frac{\ln{|W|}}{2}\right)-\ln{|W|}\notag\right.\\
	&\left.\quad+\frac{P_{0+}}{\nu V_{0+}}\frac{W\ln{|W|}+W-1}{1+W}\xi\right\}
	+ b^2V_{0+}\nu\frac{d V_0}{d z}\Big|_{+}\left(3+2W+\frac{2W\ln{|W|}}{1+W}\right).
	\label{eqn:heat_fuel_compression_zone}
\end{align}
The integral constant $\check{\theta}_{1c}$ is determined by taking $\xi\to+\infty$ in (\ref{eqn:heat_fuel_compression_zone}), with the help of matching condition (\ref{matching_conditions_all_compression}) while being careful about $dT_0/dz|_+=d^2T_0/dz^2|_+$ which is obtained by taking $\xi\to+\infty$ in (\ref{eqn:heat_fuel_compression_zone_2}), as
\begin{align}
	\check{\theta}_{1c} &= \frac{1+\alpha}{\alpha}P_{0+}\check{v}_{1+}-{T}_{1+}\left(\frac{\gamma}{\gamma-1}+\frac{1}{\alpha}-\ln{|\alpha|}\right) + \frac{\gamma}{\gamma-1}\frac{d T_1}{d z}\Big|_{+}\notag\\
	&\quad-P_{1+}V_{0+}\left(1+\alpha-\frac{1}{\alpha}+2\ln{|\alpha|}\right)+b^2{V}_{0+}^2\frac{1-B}{4}\left(1+6\alpha+\alpha^2+4\ln{|\alpha|}\right)\notag\\
	&\quad+b^2{V}_{0+}^2\frac{\gamma}{4}(\alpha-1)(\alpha+3)+ b^2\frac{V_{0+}}{P_{0+}}\nu\frac{d T_0}{d z}\Big|_{+}\left((1-\alpha)(1-\ln{|\alpha|})+\frac{(\ln{|\alpha|})^2}{2}\right)\notag\\
	&\quad+ b^2V_{0+}\nu\frac{d V_0}{d z}\Big|_{+}(3+\alpha+2\ln{|\alpha|})
	,\label{eqn:heat_fuel_compression_zone_c}
\end{align}
where we note that $\lim_{\xi\to+\infty}\xi W=0$ and $\lim_{\xi\to+\infty}\xi W\ln{|W|}=0$, which are readily confirmed by l'H\^opital's theorem.
\begin{remark}
	The terms $\ln |W|+\xi{P}_{0+}/(\nu V_{0+})$ in (\ref{eqn:heat_fuel_compression_zone}) converge to $\alpha+\ln{|\alpha|}$ as $\xi\to+\infty$.
	In fact, the following relation holds.
	\begin{align}
		\ln{|W|}+\frac{{P}_{0+}}{\nu V_{0+}}\xi=-W+\ln|\alpha|+\alpha.\label{rel:W}
	\end{align}
\end{remark}
The relation (\ref{rel:W}) is obtained by taking logarithm of both sides of the following equation.
\begin{align}
	W\mathrm{e}^W = \alpha\exp\left(-\frac{1}{\nu}\frac{{P}_{0+}}{{V}_{0+}}\xi+\alpha\right),
\end{align}
which is the definition of Lambert $W$ function for the solution (\ref{sol:v0_compression}) \cite{Wada21apr}.

\subsection{Reaction Zone $(-\infty<\eta<0)$}
\label{subsec:Reaction zone}
\subsubsection{$O(\epsilon^0)$ Solutions in Reaction Zone}
\label{subsubsec:O(epsilon0) solutions in reaction zone}
Collecting $O(\epsilon^0)$ terms, governing equations (\ref{mass_flux_unity}), (\ref{eqn:momentum}) and (\ref{eqn:state}) become
\begin{align}
	\hat{\rho}_0\hat{v}_0 &= 1,\label{eqn:mass_reaction_zone_0}\\
	0 &= -\frac{d \hat{p}_0}{d \eta} + b^2\nu\frac{d^2 \hat{v}_0}{d \eta^2},\label{eqn:momentum_reaction_zone_0}\\
	\hat{p}_0 &= \hat{\rho}_0\hat{\theta}_0.\label{eqn:state_reaction_zone_0}
\end{align}
In the same procedure as Section \ref{subsubsec:O(epsilon0) solutions in compression zone}, we have the following system of differential equations in the reaction zone.
\begin{align}
	\frac{\hat{v}_{0}}{\hat{v}_{0}-{V}_{0-}}\frac{d \hat{v}_{0}}{d \eta} &= -\frac{{P}_{0-}}{b^2\nu},\label{eqn:nonlinear_v_0_negative}\\
	\hat{p}_{0} &= b^2\nu\frac{d \hat{v}_{0}}{d \eta} + {P}_{0-}.\label{pressure_0_integral_negative}
\end{align}
Because $\hat{v}_{0}$ should be bounded in space, we find that only constant solutions are admissible for (\ref{eqn:nonlinear_v_0_negative}) and (\ref{pressure_0_integral_negative}) as follows.
\begin{align}
	\hat{v}_{0} = {V}_{0-},\quad \hat{p}_{0} = {P}_{0-}.\label{sol:v0_p0_negative}
\end{align}
Due to (\ref{eqn:heat_fuel_reaction_zone_0}), (\ref{eqn:state_reaction_zone_0}) and (\ref{sol:v0_p0_negative}), the density at $O(\epsilon^0)$ is also constant.
\begin{align}
	\hat{\rho}_{0} = {P}_{0-}/T_b.\label{sol:rho_0_negative}
\end{align}

\subsubsection{$O(\epsilon)$ Solutions in Reaction Zone}
\label{subsubsec:O(epsilon) solutions in reaction zone}
At $O(\epsilon)$, the governing equations (\ref{mass_flux_unity})-(\ref{eqn:state}) are written down, while being careful about the constancy of leading-order solutions (\ref{sol:v0_p0_negative}), as
\begin{align}
	0 &= \hat{\rho}_{1}\hat{v}_{0}+\hat{\rho}_{0}\hat{v}_{1},\label{eqn:mass_reaction_zone_1}\\
	0 &= -\frac{d \hat{p}_1}{d \eta} + b^2\nu\frac{d^2 \hat{v}_1}{d \eta^2},\label{eqn:momentum_reaction_zone_1}\\
	0 &= \frac{d^2 \hat{\theta}_1}{d \eta^2} + q\hat{\Lambda}_0\hat{\rho}_0\hat{c}_1\mathrm{e}^{\hat{\theta}_1},\label{eqn:heat_reaction_zone_1}\\
	0 &= \frac{1}{Le}\frac{d^2 \hat{c}_1}{d \eta^2} - \hat{\Lambda}_0\hat{\rho}_0\hat{c}_1\mathrm{e}^{\hat{\theta}_1},\label{eqn:fuel_reaction_zone_1}\\
	\hat{p}_1 &= \hat{\rho}_1\hat{\theta}_0 + \hat{\rho}_0\hat{\theta}_1.\label{eqn:state_reaction_zone_1}
\end{align}
At first, we calculate the temperature gradient at $O(\epsilon)$, $d\hat{\theta}_1/d\eta$.
To that end, we combine (\ref{eqn:heat_reaction_zone_1}) and (\ref{eqn:fuel_reaction_zone_1}) to eliminate the reaction terms.
\begin{align}
	0 = \frac{d^2}{d \eta^2}\left(\hat{\theta}_{1}+\frac{q}{Le}\hat{c}_{1}\right).\label{eqn:heat_fuel_reaction_1_negative}
\end{align}
Integrating (\ref{eqn:heat_fuel_reaction_1_negative}), with the help of (\ref{matching_conditions_d_reaction}), we get
\begin{align}
	\frac{d}{d \eta}\left(\hat{\theta}_{1}+\frac{q}{Le}\hat{c}_{1}\right) = \frac{d}{d z}\left({T}_{0}+\frac{q}{Le}{Y}_{0}\right)\Big|_{-}.\label{eqn:heat_fuel_reaction_1_negative_integrate}
\end{align}
Accounting for (\ref{matching_conditions_reaction}), further integration of (\ref{eqn:heat_fuel_reaction_1_negative_integrate}) leads to
\begin{align}
	\hat{\theta}_{1}+\frac{q}{Le}\hat{c}_{1} = {T}_{1-}+\frac{q}{Le}{Y}_{1-} + \eta\frac{d}{d z}\left({T}_{0}+\frac{q}{Le}{Y}_{0}\right)\Big|_{-}.\label{eqn:heat_fuel_reaction_1_negative_integrate2}
\end{align}
By use of (\ref{eqn:heat_fuel_reaction_1_negative_integrate2}), the heat-conduction equation (\ref{eqn:heat_reaction_zone_1}) is rewritten as
\begin{align}
	0 = \frac{d^2 \hat{\theta}_{1}}{d \eta^2}+Le\hat{\Lambda}_0\hat{\rho}_0\left({T}_{1-}+\frac{q}{Le}{Y}_{1-}-\hat{\theta}_{1}+\eta\frac{d}{d z}\left({T}_{0}+\frac{q}{Le}{Y}_{0}\right)\Big|_{-}\right)\mathrm{e}^{\hat{\theta}_1}.\label{eqn:heat_reaction_1_negative2}
\end{align}
Multiplying $2d\hat{\theta}_1/d\eta$ to (\ref{eqn:heat_reaction_1_negative2}), we obtain
\begin{align}
	&0 = \frac{d }{d \eta}\left(\frac{d \hat{\theta}_{1}}{d \eta}\right)^2-2Le\hat{\Lambda}_0\hat{\rho}_0\frac{d}{d z}\left({T}_{0}+\frac{q}{Le}{Y}_{0}\right)\Big|_{-}\mathrm{e}^{\hat{\theta}_1}\notag\\
	&+2Le\hat{\Lambda}_0\hat{\rho}_0\frac{d }{d \eta}\left\{\left(1+{T}_{1-}+\frac{q}{Le}{Y}_{1-}-\hat{\theta}_{1}+\eta\frac{d}{d z}\left({T}_{0}+\frac{q}{Le}{Y}_{0}\right)\Big|_{-}\right)\mathrm{e}^{\hat{\theta}_1}\right\}.\label{eqn:heat_reaction_1_negative3}
\end{align}
Integrating (\ref{eqn:heat_reaction_1_negative3}), accounting for matching conditions (\ref{matching_conditions_reaction}) and (\ref{matching_conditions_d_reaction}), we obtain
\begin{align}
	&\left(\frac{d \hat{\theta}_{1}}{d \eta}\right)^2 = \left(\frac{d {T}_{0}}{d z}\Big|_{-}\right)^2+2Le\hat{\Lambda}_0\hat{\rho}_0\frac{d}{d z}\left({T}_{0}+\frac{q}{Le}{Y}_{0}\right)\Big|_{-}\int_{-\infty}^{\eta}\mathrm{e}^{\hat{\theta}_1}d\eta\notag\\
	&-2Le\hat{\Lambda}_0\hat{\rho}_0\left(1+{T}_{1-}+\frac{q}{Le}{Y}_{1-}-\hat{\theta}_{1}+\eta\frac{d}{d z}\left({T}_{0}+\frac{q}{Le}{Y}_{0}\right)\Big|_{-}\right)\mathrm{e}^{\hat{\theta}_1},\label{eqn:heat_reaction_1_negative3_integrate}
\end{align}
where we note $\hat{\theta}_1|_{-\infty}=T_{1-}+\eta dT_{0}/dz|_-\to-\infty$ as $\eta\to-\infty$, because it is natural to consider that there exists temperature gradient on the unburned side of reaction front, $dT_{0}/dz|_-\neq0$.
Taking $\eta=0-$ in (\ref{eqn:heat_reaction_1_negative3_integrate}), with $\hat{\theta}_{1-}={T}_{1-}+q{Y}_{1-}/Le$ due to (\ref{continuity_conditions}), (\ref{sol:c_compression}) and (\ref{eqn:heat_fuel_reaction_1_negative_integrate2}), we get
\begin{align}
	0&=\left(\frac{\gamma-1}{\gamma}T_b\alpha+\frac{d {T}_{0}}{d z}\Big|_{+}\right)^2-\left(\frac{d {T}_{0}}{d z}\Big|_{-}\right)^2\notag\\
	&\quad+2Le\hat{\Lambda}_0\hat{\rho}_0\left(\mathrm{e}^{\hat{\theta}_{1-}}-\frac{d}{d z}\left({T}_{0}+\frac{q}{Le}{Y}_{0}\right)\Big|_{-}I\right),\label{eqn:heat_reaction_1_negative3_integrate_value}
\end{align}
with
\begin{align}
	I = \int_{-\infty}^{0-}\mathrm{e}^{\hat{\theta}_1}d\eta.\label{def:I}
\end{align}
The first term in (\ref{eqn:heat_reaction_1_negative3_integrate_value}) is due to (\ref{assm:heat_fuel_derivative}), (\ref{t_0_compression}), (\ref{eqn:heat_reaction_1_positive_integrate}) and (\ref{def:alpha}).
By substituting outer solutions calculated in Section \ref{sec:Outer regions} into (\ref{eqn:heat_reaction_1_negative3_integrate_value}), the leading order term of burning-rate eigenvalue $\hat{\Lambda}_0$ is calculated in Section \ref{sec:MAIN RESULTS}.

Next, we calculate $\hat{v}_1$ and $\hat{p}_1$.
Once $\hat{v}_1$ is obtained, $\hat{\rho}_1$ is calculated from (\ref{eqn:mass_reaction_zone_1}) as
\begin{align}
	\hat{\rho}_{1} = -\frac{\hat{\rho}_{0}}{\hat{v}_{0}}\hat{v}_{1}.\label{sol:rho_1_reaction}
\end{align}
Accounting for $dP_0/dz|_-=0$, which is found by taking $\eta\to-\infty$ in (\ref{eqn:momentum_reaction_zone_1}) with the help of (\ref{matching_conditions_d_reaction_2}), the integration of (\ref{eqn:momentum_reaction_zone_1}) tells us $\hat{p}_{1}$ is calculated by
\begin{align}
	\hat{p}_{1} = b^2\nu\frac{d\hat{v}_{1}}{d\eta}+P_{1-}-b^2\nu\frac{d V_{0}}{d z}\Big|_-,\label{int:p_1_reaction}
\end{align}
where (\ref{matching_conditions_reaction}) and (\ref{matching_conditions_d_reaction}) are employed.
Differentiating (\ref{eqn:state_reaction_zone_1}) with respect to $\eta$, by use of (\ref{eqn:state_reaction_zone_0}) and (\ref{sol:rho_1_reaction}), we have
\begin{align}
	\frac{d\hat{p}_{1}}{d\eta} = -\frac{\hat{p}_{0}}{\hat{v}_{0}}\frac{d\hat{v}_{1}}{d\eta}+\hat{\rho}_{0}\frac{d \hat{\theta}_{1}}{d \eta}.\label{dif:eqn_state_reaction}
\end{align}
Substituting (\ref{dif:eqn_state_reaction}) into (\ref{eqn:momentum_reaction_zone_1}), we obtain the following differential equation for $\hat{v}_1$.
\begin{align}
	\frac{d^2\hat{v}_{1}}{d\eta^2} + \frac{1}{b^2\nu}\frac{\hat{p}_{0}}{\hat{v}_{0}}\frac{d\hat{v}_{1}}{d\eta}=\frac{\hat{\rho}_{0}}{b^2\nu}\frac{d \hat{\theta}_{1}}{d \eta}.\label{eqn:v_1_reaction}
\end{align}
Putting $\lambda=\hat{p}_{0}/(b^2\nu\hat{v}_{0})$, we get
\begin{align}
	\hat{v}_{1}=V_{1-} - \frac{T_{1-}}{P_{0-}} + \frac{1}{\lambda P_{0-}}\frac{d T_{0}}{d z}\Big|_- + \frac{\hat{\theta}_{1}}{P_{0-}}-\frac{\mathrm{e}^{-\lambda\eta}}{P_{0-}}\int_{-\infty}^{\eta}\mathrm{e}^{\lambda\eta}\frac{d \hat{\theta}_{1}}{d \eta}d\eta,\label{sol:v_1_reaction}
\end{align}
where we required $\hat{v}_{1}$ is bounded in space.
In determining integral constant, we note matching relation $dV_0/dz|_-=(\hat{\rho}_{0}\hat{v}_{0}/\hat{p}_{0})dT_0/dz|_-$, which is obtained by taking $\eta\to-\infty$ in (\ref{dif:eqn_state_reaction}), and the limit $\lim_{\eta\to-\infty}\mathrm{e}^{-\lambda\eta}\int_{-\infty}^{\eta}\mathrm{e}^{\lambda\eta}(d \hat{\theta}_{1}/d \eta) d\eta=\lambda^{-1}dT_0/dz|_-$, which is confirmed by l'H\^opital's theorem.
Then, from (\ref{int:p_1_reaction}) and (\ref{sol:v_1_reaction}), $\hat{p}_{1}$ is calculated as
\begin{align}
	\hat{p}_{1} = P_{1-}-b^2\nu\frac{d V_{0}}{d z}\Big|_-+\frac{\mathrm{e}^{-\lambda\eta}}{V_{0-}}\int_{-\infty}^{\eta}\mathrm{e}^{\lambda\eta}\frac{d \hat{\theta}_{1}}{d \eta}d\eta.\label{sol:p_1_reaction}
\end{align}

\subsubsection{Heat-Conduction and Mass-Fraction Equations at $O(\epsilon^2)$ in Reaction Zone}
\label{subsubsec:O(epsilon2) solutions in reaction zone}
At $O(\epsilon^2)$, with the help of (\ref{sol:v0_p0_negative}), we consider the combination of heat-conduction equation (\ref{eqn:heat}) and mass-fraction equation (\ref{eqn:fuel}), which is written as
\begin{align}
	\frac{d }{d \eta}(\hat{\theta}_1+q\hat{c}_1) = \frac{d^2 }{d \eta^2}\left(\hat{\theta}_2+\frac{q}{Le}\hat{c}_2\right) + \frac{\gamma - 1}{\gamma}\hat{v}_0\frac{d \hat{p}_1}{d \eta}.\label{eqn:heat_fuel_reaction_zone_2}
\end{align}
The integration of (\ref{eqn:heat_fuel_reaction_zone_2}) leads to
\begin{align}
	\hat{\theta}_1+q\hat{c}_1 &= \frac{d }{d \eta}\left(\hat{\theta}_2+\frac{q}{Le}\hat{c}_2\right) +\frac{\gamma - 1}{\gamma}\hat{v}_0\hat{p}_1+{T}_{1-}+q{Y}_{1-}\notag\\
	&\quad- \frac{d }{d z}\left({T}_1+\frac{q}{Le}{Y}_1\right)\Big|_--\frac{\gamma - 1}{\gamma}{V}_{0-}{P}_{1-},\label{eqn:heat_fuel_reaction_zone}
\end{align}
where the matching conditions (\ref{matching_conditions_reaction}) and (\ref{matching_conditions_d2_reaction}) are used to determine the integral constant in the limit of $\eta\to-\infty$, while being careful about the matching relation $d(T_0+qY_0)/d z|_-=d^2(T_0+qY_0/Le)/d z^2|_-$, which is obtained by taking $\eta\to-\infty$ in (\ref{eqn:heat_fuel_reaction_zone_2}) by use of (\ref{matching_conditions_d_reaction_2}) and $dP_0/d z|_-=0$.

\subsection{Boundary Conditions Across Reaction and Compression Zones}
\label{subsec:Boundary conditions across inner regions}
We summarize the boundary conditions across the reaction and compression zones, which are obtained subject to assumptions (\ref{continuity_conditions}) and (\ref{assm:heat_fuel_derivative}).
From (\ref{eqn:heat_fuel_reaction_zone_0}), (\ref{sol:c_compression}), (\ref{t_0_compression}), (\ref{eqn:heat_reaction_1_positive_integrate}), (\ref{sol:theta1_compression}), (\ref{eqn:heat_fuel_reaction_1_negative_integrate}), (\ref{eqn:v1_compression_zone_1}), (\ref{eqn:heat_fuel_reaction_1_negative_integrate2}) and (\ref{int:p_1_reaction}), by taking $\eta=0-$ or $\xi=0+$, we find
\begin{align}
	[{T}_{0}] = [{Y}_{0}]&=0,\label{bdc:heat_mass0}\\
	\left[\frac{d }{d z}\left({T}_{0}+\frac{q}{Le}{Y}_{0}\right)\right]&=-\frac{\gamma-1}{\gamma}T_b\alpha,\label{bdc:heat_mass_derivative0}\\
	\left[{T}_{1} +\frac{q}{Le}{Y}_{1}\right]&=b^2\frac{\gamma-1}{\gamma}\nu{V}_{0+}^2\alpha\left(1+\frac{\alpha}{2}\right),\label{bdc:heat_mass1}\\
	[P_1]&=b^2\left(\nu\left[\frac{d V_0}{d z}\right]+\gamma\alpha\right),\label{bdc:momentum1}
\end{align}
where the brackets $[\;]$ means the difference of solutions evaluated at the burned-side edge of compression zone and the unburned-side edge of reaction zone, $[\Phi(z)]=\Phi_+-\Phi_-$.
The constant $\alpha$ is defined by (\ref{def:alpha}) and now found to be $\alpha=V_{0-}/V_{0+}-1$.

From (\ref{continuity_conditions}), (\ref{sol:c_compression}), (\ref{eqn:heat_fuel_reaction_1_negative_integrate2}) and (\ref{sol:v_1_reaction}), we find
\begin{align}
	\check{v}_{1+}=\hat{v}_{1-}=V_{1-} + \frac{q}{Le}\frac{Y_{1-}}{P_{0-}} + \frac{1}{\lambda P_{0-}}\frac{d T_{0}}{d z}\Big|_- -\frac{1}{P_{0-}}\int_{-\infty}^{0-}\mathrm{e}^{\lambda\eta}\frac{d \hat{\theta}_{1}}{d \eta}d\eta,\label{v_1+}
\end{align}

Taking $\xi=0+$ in (\ref{eqn:heat_fuel_compression_zone}) and $\eta=0-$ in (\ref{eqn:heat_fuel_reaction_zone}), by use of (\ref{continuity_conditions}) and (\ref{assm:heat_fuel_derivative}), we obtain
\begin{align}
	0 &= \frac{\gamma}{\gamma - 1}\left\{\left[\frac{d }{d z}\left(T_1+\frac{q}{Le}Y_1\right)\right]+q\left(1-\frac{1}{Le}\right)Y_{1-}\right\}\notag\\
	&\quad+ (1+\alpha)[V_{0}P_{1}] + \alpha T_{1+} + \frac{\alpha}{\lambda}\frac{d T_{0}}{d z}\Big|_- - \alpha\int_{-\infty}^{0-}\mathrm{e}^{\lambda\eta}\frac{d \hat{\theta}_{1}}{d \eta}d\eta\notag\\
	&\quad-b^2V_{0+}^2\alpha\left\{\frac{\gamma}{2}\alpha+\nu\left(1+\frac{\gamma-1}{\gamma}\alpha\right)\left(1+\frac{\alpha}{2}\right)+\frac{\nu}{V_{0+}}\frac{d V_{0}}{d z}\Big|_+\right\}.\label{bdc:heat_mass_2}
\end{align}

\section{OUTER REGIONS}
\label{sec:Outer regions}
In the previous section, we have obtained the inner solutions in the reaction and compression zones and the boundary conditions across them.
In this section, we seek for the solutions in the outer regions, which consist of the unburned and burned zones.
Now, the governing equations (\ref{eqn:momentum})-(\ref{eqn:fuel}) in the outer regions ($z<0,z>0$) are written, by use of (\ref{def:b}), as
\begin{align}
	b^2\gamma \frac{d V}{d z} &= -\frac{1}{\epsilon}\frac{d P}{d z} + b^2\nu\frac{d^2 V}{d z^2},\label{eqn:momentum_outer}\\
	\frac{d T}{d z} &= \frac{d^2 T}{d z^2} + \frac{\gamma - 1}{\gamma}\left(V\frac{d P}{d z}+\epsilon b^2\nu\left(\frac{d V}{d z}\right)^2\right),\label{eqn:heat_outer}\\
	\frac{d Y}{d z} &= \frac{1}{Le}\frac{d^2 Y}{d z^2},\label{eqn:fuel_outer}
\end{align}
where the reaction term $Q$ is omitted from (\ref{eqn:heat_outer}) and (\ref{eqn:fuel_outer}) due to the large-activation-energy limit and $T<T_b$ for $z<0$ and due to $Y=0$ for $z>0$.

\subsection{Outer Solutions}
\label{subsec:Outer solutions}
We seek for solutions of governing equations (\ref{mass_flux_unity}), (\ref{eqn:state}), (\ref{eqn:momentum_outer})-(\ref{eqn:fuel_outer}) in the form of asymptotic expansions with respect to $\epsilon$ as introduced by (\ref{epsilon_expansions_outer}).
From (\ref{eqn:momentum_outer}), with the help of (\ref{bdc:far_field}), we find that the pressure at $O(\epsilon^0)$ is constant.
\begin{align}
	{P}_0=
	\left\{
	\begin{array}{lll}
		1  & (z<0)\\
		{P}_{0+} & (z>0)
	\end{array}
	\right..\label{P0}
\end{align}
Taking account of (\ref{P0}), the solutions of other quantities at $O(\epsilon^0)$ and pressure at $O(\epsilon)$ are obtained by solving the following equations.
\begin{align}
	{R}_0{V}_0&=1,\label{eqn:R0V0}\\
	b^2\gamma\frac{d {V}_0}{d z} &= -\frac{d {P}_1}{d z} + b^2\nu\frac{d^2 {V}_0}{d z^2},\label{eqn:P2}\\
	\frac{d {T}_0}{d z} &= \frac{d^2 {T}_0}{d z^2},\label{eqn:T0}\\
	\frac{d {Y}_0}{d z} &= \frac{1}{Le}\frac{d^2 {Y}_0}{d z^2},\label{eqn:Y0}\\
	{P}_0&={R}_0{T}_0.\label{eqn:P0R0T0}
\end{align}
Boundary conditions (\ref{bdc:far_field}), (\ref{assm:Y=0}), (\ref{bdc:heat_mass0}) and (\ref{bdc:momentum1}) are imposed on (\ref{eqn:R0V0})-(\ref{eqn:P0R0T0}).
Then, the solutions in the outer regions are given as follows.
\begin{align}
	{T}_0&=
	\left\{
	\begin{array}{lll}
		1 + (T_b-1)\mathrm{e}^{z} & (z<0)\\
		T_b & (z>0)
	\end{array}
	\right.,\quad
	{Y}_0=
	\left\{
	\begin{array}{lll}
		1 - \mathrm{e}^{Lez} & (z<0)\\
		0 & (z>0)
	\end{array}
	\right.,\label{sol:T0Y0}\\
	{V}_0&=
	\left\{
	\begin{array}{lll}
		{T}_0 & (z< 0)\\
		V_{0+} & (z>0)
	\end{array}
	\right.,\quad\quad
	{R}_0= {V}_0^{-1},\label{sol:V0R0}\\
	{P}_0&=
	\left\{
	\begin{array}{lll}
		1 & (z< 0)\\
		T_bV_{0+}^{-1} & (z>0)
	\end{array}
	\right.,\quad
	{P}_1=
	\left\{
	\begin{array}{lll}
		b^2(\nu-\gamma)(T_b-1)\mathrm{e}^{z}& (z<0)\\
		b^2\gamma(1+\alpha-T_b)& (z>0)
	\end{array}
	\right.,\label{sol:P0P1}
\end{align}
We find that the following relation holds due to (\ref{bdc:heat_mass_derivative0}) and (\ref{sol:T0Y0}).
\begin{align}
	\frac{\gamma-1}{\gamma}T_b\alpha=T_b-(1+q)=T_b-T_{ad}.\label{rel:Tb_Tad}
\end{align}
Then, by use of (\ref{def:alpha}), we get
\begin{align}
	V_{0+} = \frac{(\gamma-1)T_b^2}{(2\gamma-1)T_b-\gamma(1+q)}.\label{def:A}
\end{align}
We note that if $T_b$ takes its adiabatic value $T_{ad}=1+q$, the adiabatic profile of a premixed flame front is recovered.
However, in the present case, its value is not restricted to adiabatic one, and the non-zero temperature gradient is described by the deviation of $T_b$ from $T_{ad}$ as will be shown in (\ref{eqn:heat_reaction_1_positive_integrate_0+}).

At $O(\epsilon)$, governing equations (\ref{eqn:heat_outer}) and (\ref{eqn:fuel_outer}) are written down as
\begin{align}
	\frac{d {T}_1}{d z} &= \frac{d^2 {T}_1}{d z^2} + \frac{\gamma - 1}{\gamma}\left({V}_0\frac{d {P}_1}{d z}+b^2\nu\left(\frac{d V_0}{d z}\right)^2\right),\label{eqn:heat_outer_1}\\
	\frac{d {Y}_1}{d z} &= \frac{1}{Le}\frac{d^2 {Y}_1}{d z^2}.\label{eqn:fuel_outer_1}
\end{align}
Solving (\ref{eqn:heat_outer_1}) and (\ref{eqn:fuel_outer_1}), by use of (\ref{bdc:far_field}) and (\ref{assm:Y=0}), we obtain
\small
\begin{align}
	{T}_1&=
	\left\{
	\begin{array}{lll}
		C_T\mathrm{e}^{z} - b^2\frac{\gamma - 1}{\gamma}(T_b-1)\big\{(\nu-\gamma)z+\frac{T_b-1}{2}(2\nu-\gamma)\mathrm{e}^{z}\big\}\mathrm{e}^{z} & (z<0)\\
		{T}_{1+} & (z>0)
	\end{array}
	\right.,\label{sol:T1}\\
	{Y}_1&=
	\left\{
	\begin{array}{lll}
		C_Y\mathrm{e}^{Lez} & (z<0)\\
		0 & (z>0)
	\end{array}
	\right.,\label{sol:Y1}
\end{align}
\normalsize
where $C_T$ and $C_Y$ are integral constants.


The value of $T_{1+}$ is calculated from (\ref{bdc:heat_mass_2}), with (\ref{sol:V0R0}), (\ref{sol:P0P1}), (\ref{sol:T1}) and (\ref{sol:Y1}) substituted, as follows.
\begin{align}
	T_{1+}&=\frac{\gamma-1}{\gamma}\frac{T_b}{T_{ad}}\left\{\frac{b^2}{2}\left(\gamma+T_b^2\frac{\gamma(\alpha^3-1)-(B-1)\alpha^2(\alpha+2)}{(1+\alpha)^2}\right)\right.\notag\\
	&\left.\quad-\alpha\int_{-\infty}^{0-}\mathrm{e}^{\lambda\eta}\frac{d\hat{\theta}_1}{d\eta}d\eta\right\},\label{T1+}
\end{align}

\section{MAIN RESULTS}
\label{sec:MAIN RESULTS}
\subsection{Formula of Laminar Flame Speed}
\label{subsec:Formula of Laminar flame speed}
The formula of laminar flame speed is obtained by calculating the burning-rate eigenvalue \cite{Friedman53,Karman57,Bush70}.
The leading-order term of burning-rate eigenvalue $\hat{\Lambda}_0$ is calculated by substituting (\ref{sol:T0Y0}) into (\ref{eqn:heat_reaction_1_negative3_integrate_value}), with the help of (\ref{rel:Tb_Tad}), as
\begin{align}
	\hat{\Lambda}_0=\frac{q}{Le\hat{\rho}_0}\left(T_b-1-\frac{q}{2}\right)\left(\mathrm{e}^{\hat{\theta}_{1-}}-(T_b-1-q)I\right)^{-1}.\label{sol:burning-rate-eigenvalue}
\end{align}
Recalling (\ref{def:epsilon}), (\ref{Lambda_rewritten}) and (\ref{sol:rho_0_negative}), the parameter dependence of laminar flame speed is obtained from (\ref{sol:burning-rate-eigenvalue}) as
\begin{align}
	\tilde{S}_L=\left(2\tilde{A}\tilde{D}_{th}Le\frac{T_b^3}{N^2}\mathrm{e}^{-N/T_b}\frac{\mathrm{e}^{\hat{\theta}_{1-}}-(T_b-1-q)I}{2q(T_b-1)-q^2}\right)^{1/2}.\label{laminar_flame_speed}
\end{align}

\subsection{Differential Equation for Temperature}
\label{subsec:Differential Equation for Temperature}
In the reaction zone, $\eta<0$, the temperature at $O(\epsilon^0)$ has a constant value $T_b$ as mentioned in (\ref{eqn:heat_fuel_reaction_zone_0}).
At the next order, the value of $\hat{\theta}_1$ is computed from (\ref{eqn:heat_reaction_1_negative2}).
Using (\ref{continuity_conditions}), (\ref{sol:c_compression}) and (\ref{eqn:heat_fuel_reaction_1_negative_integrate2}) as $\eta=0-$ and (\ref{sol:T0Y0}), (\ref{sol:burning-rate-eigenvalue}), we transform (\ref{eqn:heat_reaction_1_negative2}) into the following form.
\begin{align}
	0 = \frac{d^2 \hat{\theta}_{1}}{d \eta^2}+q\left(T_b-1-\frac{q}{2}\right)\frac{\hat{\theta}_{1-}-\hat{\theta}_{1}+\eta (T_b-1-q)}{\mathrm{e}^{\hat{\theta}_{1-}}-(T_b-1-q)I}\mathrm{e}^{\hat{\theta}_1},\label{eqn:heat_reaction_1_negative2_trans}
\end{align}
subject to the boundary conditions
\begin{align}
	\frac{d \hat{\theta}_{1}}{d \eta}\Big|_{-} &= T_b - T_{ad},\label{eqn:heat_reaction_1_positive_integrate_0+}\\
	\hat{\theta}_{1-} &= {T}_{1+} -\frac{1}{2}{M\hspace{-1.5pt}a}^2\frac{N}{T_b^2}\nu{V}_{0+}^2\left(1-\frac{1+q}{T_b}\right)\left(1+\frac{T_b}{V_{0+}}\right).\label{eqn:heat_reaction_1_positive_integrate2_0+}
\end{align}
The first condition (\ref{eqn:heat_reaction_1_positive_integrate_0+}) is obtained by taking $\xi=0+$ in (\ref{eqn:heat_reaction_1_positive_integrate}), by use of (\ref{assm:heat_fuel_derivative}), (\ref{eqn:heat_fuel_reaction_zone_0}), (\ref{sol:T0Y0}) and (\ref{rel:Tb_Tad}).
From (\ref{eqn:heat_reaction_1_positive_integrate_0+}), we find that the temperature gradient at the burned-side edge of a flame front is simply described by the deviation of flame temperature from its adiabatic value.
The second condition (\ref{eqn:heat_reaction_1_positive_integrate2_0+}) is obtained by taking $\xi=0+$ in (\ref{sol:theta1_compression}), with the help of (\ref{def:epsilon}), (\ref{def:b}) and (\ref{rel:Tb_Tad}).

In the compression zone, $\xi>0$, the asymptotic solution of temperature is obtained from (\ref{def:epsilon}), (\ref{def:b}), (\ref{t_0_compression}), (\ref{sol:theta1_compression}) and (\ref{sol:T0Y0}) as
\begin{align}
	T = T_b + \frac{T_b^2}{N}{T}_{1+}-{M\hspace{-1.5pt}a}^2\frac{\gamma-1}{\gamma}\nu{V}_{0+}^2W\left(1+\frac{W}{2}\right)+O(N^{-2}).\label{eqn:heat_reaction_1_positive_integrate2_trans}
\end{align}

In the subsequent subsections, by use of (\ref{sol:burning-rate-eigenvalue}), the flow properties are investigated for adiabatic, inviscid and non-adiabatic cases.

\subsection{Adiabatic case}
\label{subsec:Adiabatic case}
In the adiabatic case, $T_b=T_{ad}=1+q$, we find $V_{0-}/V_{0+}-1=\alpha=0$ due to (\ref{rel:Tb_Tad}).
Then, from (\ref{T1+}), we find that the temperature on the burned side is decreased by the compressibility effect brought by $O({M\hspace{-1.5pt}a}^2)$ term, because $T_{ad}>1$ for $q>0$.
\begin{align}
	T_{1+}= b^2\frac{\gamma-1}{2}(1-T_{ad}^2).\label{T1+_ad}
\end{align}
Now, outer solutions (\ref{sol:T0Y0})-(\ref{sol:P0P1}), (\ref{sol:T1}) and (\ref{sol:Y1}) read
\begin{align}
	{T}_0&=
	\left\{
	\begin{array}{lll}
		1 + q\mathrm{e}^{z} & (z<0)\\
		1+q & (z>0)
	\end{array}
	\right.,\quad
	{Y}_0=
	\left\{
	\begin{array}{lll}
		1 - \mathrm{e}^{Lez} & (z<0)\\
		0 & (z>0)
	\end{array}
	\right.,\label{sol:T0Y0_ad}\\
	{V}_0&=	{T}_0,\quad
	{R}_0= {V}_0^{-1},\quad
	{P}_0=1\quad(z\gtrless0),\label{sol:V0R0P_ad}\\
	{P}_1&=
	\left\{
	\begin{array}{lll}
		b^2(\nu-\gamma)q\mathrm{e}^{z}& (z<0)\\
		-b^2\gamma q& (z>0)
	\end{array}
	\right.,\quad
	{Y}_1=
	\left\{
	\begin{array}{lll}
		C_Y\mathrm{e}^{Lez} & (z<0)\\
		0 & (z>0)
	\end{array}
	\right.,\label{sol:P1Y1_ad}\\
	{T}_1&=
	\left\{
	\begin{array}{lll}
		C_T\mathrm{e}^{z} - b^2\frac{\gamma - 1}{\gamma}q\left((\nu-\gamma)z+\frac{q}{2}(2\nu-\gamma)\mathrm{e}^{z}\right)\mathrm{e}^{z} & (z<0)\\
		b^2\frac{\gamma-1}{2}(1-T_{ad}^2) & (z>0)
	\end{array}
	\right.,\label{sol:T1_ad}
\end{align}
and inner solutions in compression zone (\ref{sol:c_compression}), (\ref{eqn:mass_compression_zone_0}), (\ref{t_0_compression}), (\ref{sol:v0_compression}) and (\ref{sol:p0_compression}) are
\begin{align}
	\check{c}_i = 0\;(i=0,1\ldots),\quad \check{\theta}_0 = \check{v}_0 = \check{\rho}_0^{-1} = T_{ad},\quad \check{p}_0 = 1,\\
	\check{\theta}_1 = b^2\frac{\gamma-1}{2}(1-T_{ad}^2),\quad \check{v}_1 = b^2q\left(1+\frac{q}{2}(1+\gamma)\right),\quad \check{p}_1 = -b^2\gamma q,
\end{align}
and those in reaction zone (\ref{eqn:heat_fuel_reaction_zone_0}), (\ref{sol:v0_p0_negative}), (\ref{sol:rho_0_negative}), (\ref{eqn:heat_fuel_reaction_1_negative_integrate2}), (\ref{sol:v_1_reaction}) and (\ref{sol:p_1_reaction}) are
\begin{align}
	\hat{c}_0 = 0,\quad \hat{\theta}_0 = \hat{v}_0 = \hat{\rho}_0^{-1} = T_{ad},\quad \hat{p}_0 = 1,\\
	\hat{c}_1 = C_Y + \frac{Le}{q}\left(C_T-b^2\frac{\gamma-1}{\gamma}\frac{q^2}{2}(2\nu-\gamma)-\hat{\theta}_1\right),\\
	\hat{v}_1 = b^2\gamma qT_{ad} + \hat{\theta}_{1}-\mathrm{e}^{-\lambda\eta}\int_{-\infty}^{\eta}\mathrm{e}^{\lambda\eta}\frac{d \hat{\theta}_{1}}{d \eta}d\eta,\\
	\hat{p}_1 = -b^2\gamma q + \frac{\mathrm{e}^{-\lambda\eta}}{T_{ad}}\int_{-\infty}^{\eta}\mathrm{e}^{\lambda\eta}\frac{d \hat{\theta}_{1}}{d \eta}d\eta,
\end{align}
The value of $\hat{\theta}_1$ is computed from the differential equation (\ref{eqn:heat_reaction_1_negative2_trans}) for the adiabatic case:
\begin{align}
0 = \frac{d^2 \hat{\theta}_{1}}{d \eta^2}-\frac{q^2}{2}\left(\hat{\theta}_{1}-T_{1+}\right)\mathrm{e}^{\hat{\theta}_1-T_{1+}}.\label{eqn:t1_ad}
\end{align}

Accounting for (\ref{rel:Tb_Tad}), (\ref{T1+}) and (\ref{eqn:heat_reaction_1_positive_integrate2_0+}), the value of laminar flame speed for the adiabatic case, $\tilde{S}_{Lad}$, is expressed as follows.
\begin{align}
	\tilde{S}_{Lad}=\left(2\tilde{A}\tilde{D}_{th}Le\frac{T_{ad}^3}{q^2N^2}\mathrm{e}^{T_{1+}-N/T_{ad}}\right)^{1/2}.\label{laminar_flame_speed_ad}
\end{align}
Taking $b\to0$ in (\ref{laminar_flame_speed_ad}), which means $T_{1+}\to0$ due to (\ref{T1+}), we recover the formula of laminar flame speed in the zero-Mach-number model \cite{Joulin98,Williams85}:
\begin{align}
	\tilde{S}_{Lzero}=\left(2\tilde{A}\tilde{D}_{th}Le\frac{T_{ad}^3}{q^2N^2}\mathrm{e}^{-N/T_{ad}}\right)^{1/2}.\label{laminar_flame_speed_zero}
\end{align}
Because it is convenient to treat non-dimensional quantities, we introduce ${M\hspace{-1.5pt}a}_{ad}=\tilde{S}_{Lad}/\tilde{c}_s$ and ${M\hspace{-1.5pt}a}_{zero}=\tilde{S}_{Lzero}/\tilde{c}_s$, which are the values in the adiabatic low-Mach-number and zero-Mach-number cases, respectively.
Then, from (\ref{laminar_flame_speed_ad}) and (\ref{laminar_flame_speed_zero}), we have
\begin{align}
	{M\hspace{-1.5pt}a}_{ad}={M\hspace{-1.5pt}a}_{zero}\exp\left(\frac{T_{1+}}{2}\right).\label{rel:Ma_ad_Ma_zero}
\end{align}
From (\ref{def:b}), (\ref{T1+_ad}) and (\ref{rel:Ma_ad_Ma_zero}), the adiabatic value of $T_{1+}$ is expressed by use of the upper branch of Lambert $W$ function as
\begin{align}
	T_{1+}=-W_0\left(\frac{{M\hspace{-1.5pt}a}_{zero}^2}{\epsilon}\frac{\gamma-1}{2}(T_{ad}^2-1)\right).\label{T1+_ad_W}
\end{align}

\subsection{Inviscid Limit $Pr\to0$}
For inviscid limit $Pr\to0$, or $\nu\to0$, inner solutions are approximated by constant values because $W\to0$ for $\nu\to0$ in the range of $0<\xi<+\infty$ as confirmed from (\ref{W=W0}):
for example, $\check{v}_{0}={V}_{0+}$ which implies ${V}_{0-}={V}_{0+}$, or $\alpha=0$, due to (\ref{continuity_conditions}) and (\ref{sol:v0_p0_negative}).
Therefore, from (\ref{rel:Tb_Tad}), the temperature on the burned side of a flame front is equal to its adiabatic value.
\begin{align}
	T_b = T_{ad} = 1+q.\label{Tb_nu=0}
\end{align}
As a consequence, if the flow is inviscid, then it is adiabatic and its solutions are given by those in Section \ref{subsec:Adiabatic case} with $\nu=0$.

\subsection{Non-adiabatic case}
In this section, we reveal the relation between the laminar flame speed and the temperature profile for a viscous compressible flow.
Because it is convenient to use ${M\hspace{-1.5pt}a}$ rather than $\tilde{S}_{L}$, we transform (\ref{laminar_flame_speed}) into the following non-dimensional form, by use of (\ref{nondimensionalization}) and (\ref{laminar_flame_speed_ad}):
\begin{align}
	{M\hspace{-1.5pt}a}={M\hspace{-1.5pt}a}_{zero}\left(q\left(\frac{T_{b}}{T_{ad}}\right)^3\mathrm{e}^{N(1/T_{ad}-1/T_b)}\frac{\mathrm{e}^{\hat{\theta}_{1-}}-(T_b-1-q)I}{2(T_b-1)-q}\right)^{1/2}.\label{rel:Mach_number}
\end{align}
Requiring $\frac{\mathrm{e}^{\hat{\theta}_{1-}}-(T_b-1-q)I}{2(T_b-1)-q}>0$, we find the value of $T_b$ is limited to
\begin{align}
	1+\frac{q}{2}<T_b<1+q+\frac{\mathrm{e}^{\hat{\theta}_{1-}}}{I}.
\end{align}
The effect of non-zero temperature gradient denoted by the deviation of flame temperature from its adiabatic value, on the laminar flame speed, or the Mach number, is investigated by solving (\ref{eqn:heat_reaction_1_negative2_trans}) and (\ref{rel:Mach_number}) subject to boundary conditions (\ref{eqn:heat_reaction_1_positive_integrate_0+}) and (\ref{eqn:heat_reaction_1_positive_integrate2_0+}).
The computation of $\hat{\theta}_{1}$ and ${M\hspace{-1.5pt}a}$ from these equations is made by iterative method:
at first, we calculate the adiabatic temperature solution by solving (\ref{eqn:t1_ad}).
This solution is, in turn, used as an initial value in studying the non-adiabatic case.
For example, by fixing some value of ${M\hspace{-1.5pt}a}$ with given parameters, we change the value of $d\hat{\theta}_{1}/d\eta|_{-}\neq0$ so as to $I$ defined by (\ref{def:I}) converges numerically, where $\hat{\theta}_{1}$ is computed from (\ref{eqn:heat_reaction_1_negative2_trans}).
Next, we check the numerical convergence of ${M\hspace{-1.5pt}a}$ from (\ref{rel:Mach_number}), and if it does not converge, then we repeat the previous step on $d\hat{\theta}_{1}/d\eta|_{-}\neq0$.
In this study, we set parameters as $q=2$, $\gamma=1.4$, $N=90$ and ${M\hspace{-1.5pt}a}_{zero}=0.1$.
We investigate the effect of non-zero temperature gradient at the burned-side edge of a flame front, $d\hat{\theta}_{1}/d\eta|_{-}\neq0$ in (\ref{eqn:heat_reaction_1_positive_integrate_0+}), due to the gas compression.
Besides, the influence of viscosity, expressed by Prandtl number $Pr$, is also revealed.

\begin{figure}[t]
	\begin{minipage}{0.45\hsize}
		\begin{center}
			\includegraphics[scale=0.4]{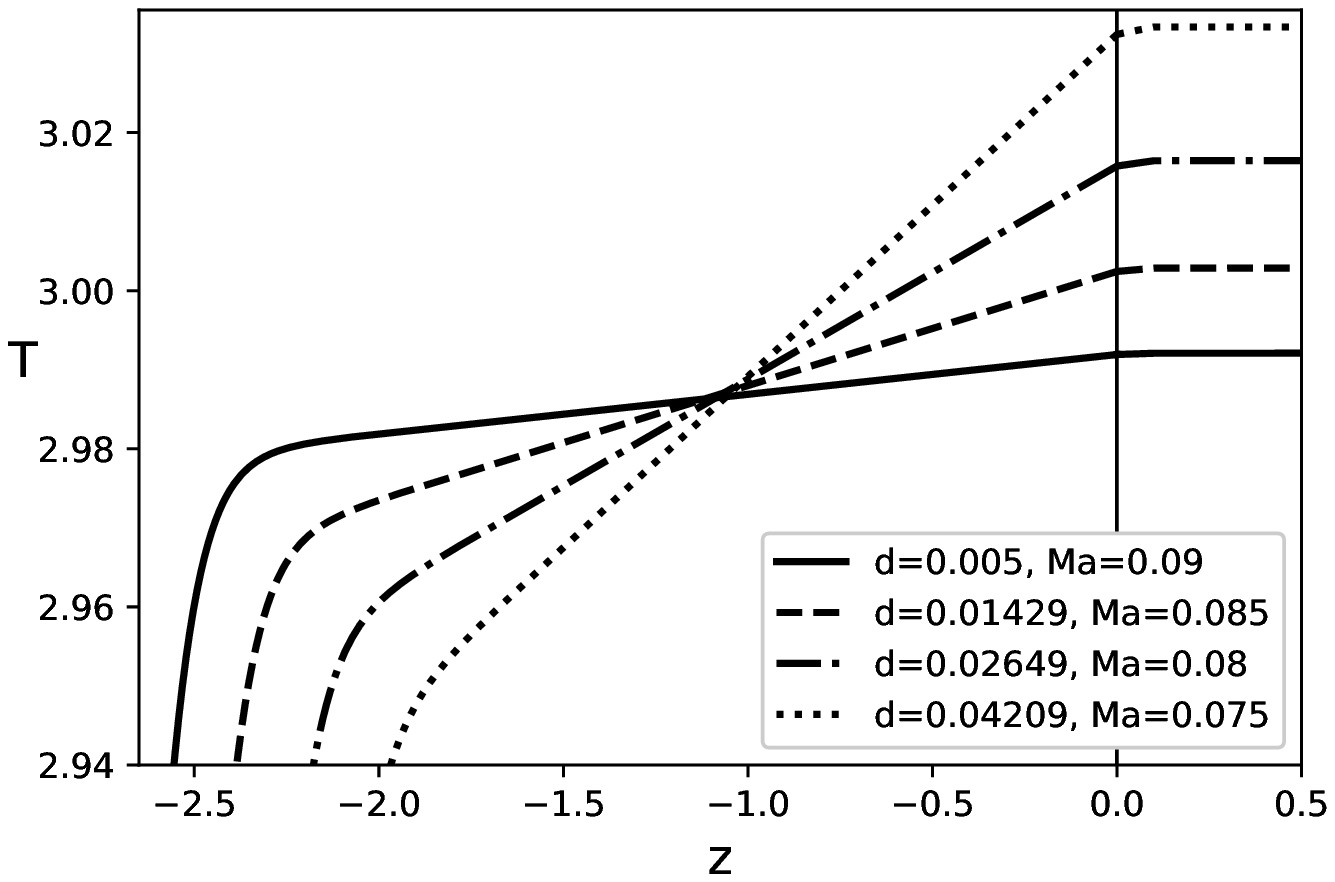}
		\end{center}
		\caption{Temperature distribution with $q=2$, $N=90$, $\gamma=1.4$, ${M\hspace{-1.5pt}a}_{zero}=0.1$ and $Pr=3/4$ for positive values of temperature gradient given in (\ref{eqn:heat_reaction_1_positive_integrate_0+}): $d=d\hat{\theta}_{1}/d\eta|_{-}$.}
		\label{fig1}
	\end{minipage}
	\hspace{1cm}
	\begin{minipage}{0.45\hsize}
		\begin{center}
			\includegraphics[scale=0.4]{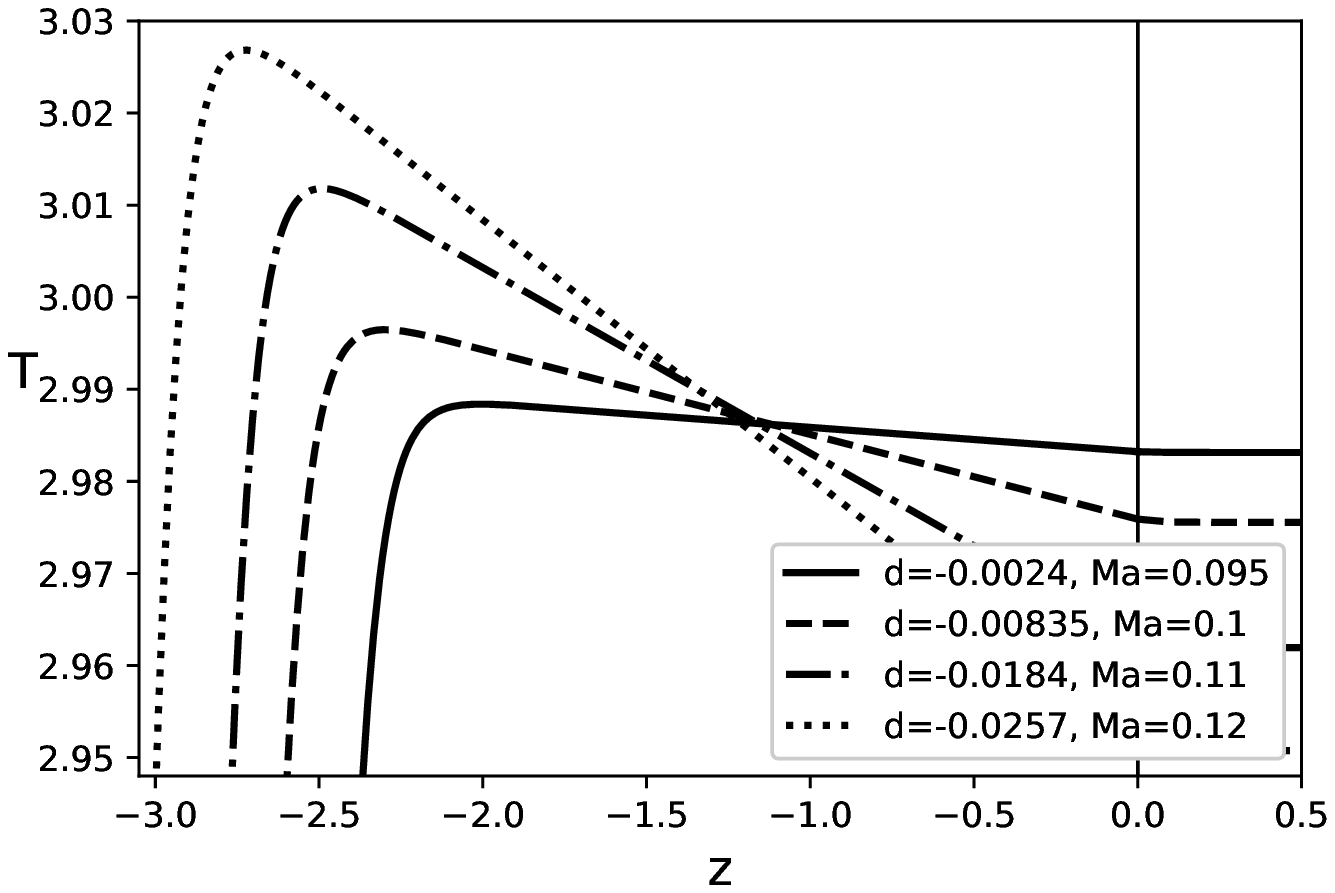}
		\end{center}
		\caption{The same as Fig. \ref{fig1} but for negative values of temperature gradient.\newline\newline}
		\label{fig2}
	\end{minipage}
\end{figure}

For the fixed value of Prandtl number $Pr=3/4$, the temperature profiles are shown in Fig. \ref{fig1} for $d\hat{\theta}_{1}/d\eta|_{-}>0$ and in Fig. \ref{fig2} for $d\hat{\theta}_{1}/d\eta|_{-}<0$.
Figure \ref{fig1} shows that the positive value of temperature gradient at the burned-side edge of a flame front brings the monotonic increase of temperature.
This behavior is similar to that in adiabatic case except for non-zero values of temperature gradient.
On the other hand, the negative gradient of temperature induces the non-monotonic profile with the maximum value of temperature achieved inside the reaction zone, as plotted in Fig. \ref{fig2}.
Such a profile is to be observed, for example, in the numerical study of thermonuclear flames based on the deflagration model \cite{Glazyrin13}, detonation model \cite{Khokhlov89,Sharpe99} or reconstruction model \cite{Townsley16}.
Our results are consistent with those obtained in previous works qualitatively.
In the present study, the asymptotic solution of temperature on the burned side, $z>0$, is described by Lamber $W$ function as obtained in (\ref{eqn:heat_reaction_1_positive_integrate2_trans}).

\begin{figure}[t]
	\begin{minipage}{0.45\hsize}
		\begin{center}
			\includegraphics[scale=0.4]{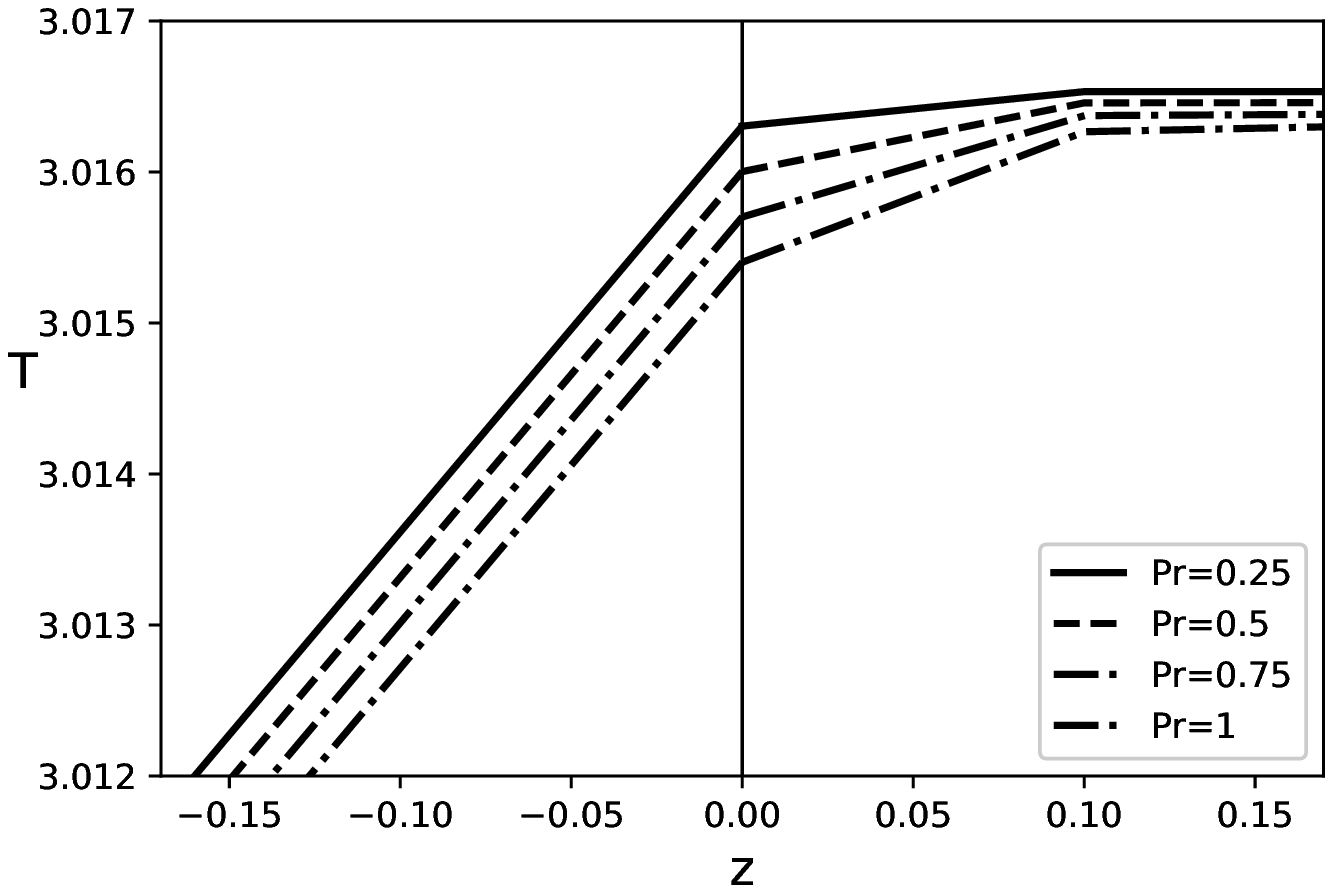}
		\end{center}
		\caption{The same as Fig. \ref{fig1} but for several values of Prandtl number with ${M\hspace{-1.5pt}a}=0.08$.}
		\label{fig3}
	\end{minipage}
	\hspace{1cm}
	\begin{minipage}{0.45\hsize}
		\begin{center}
			\includegraphics[scale=0.4]{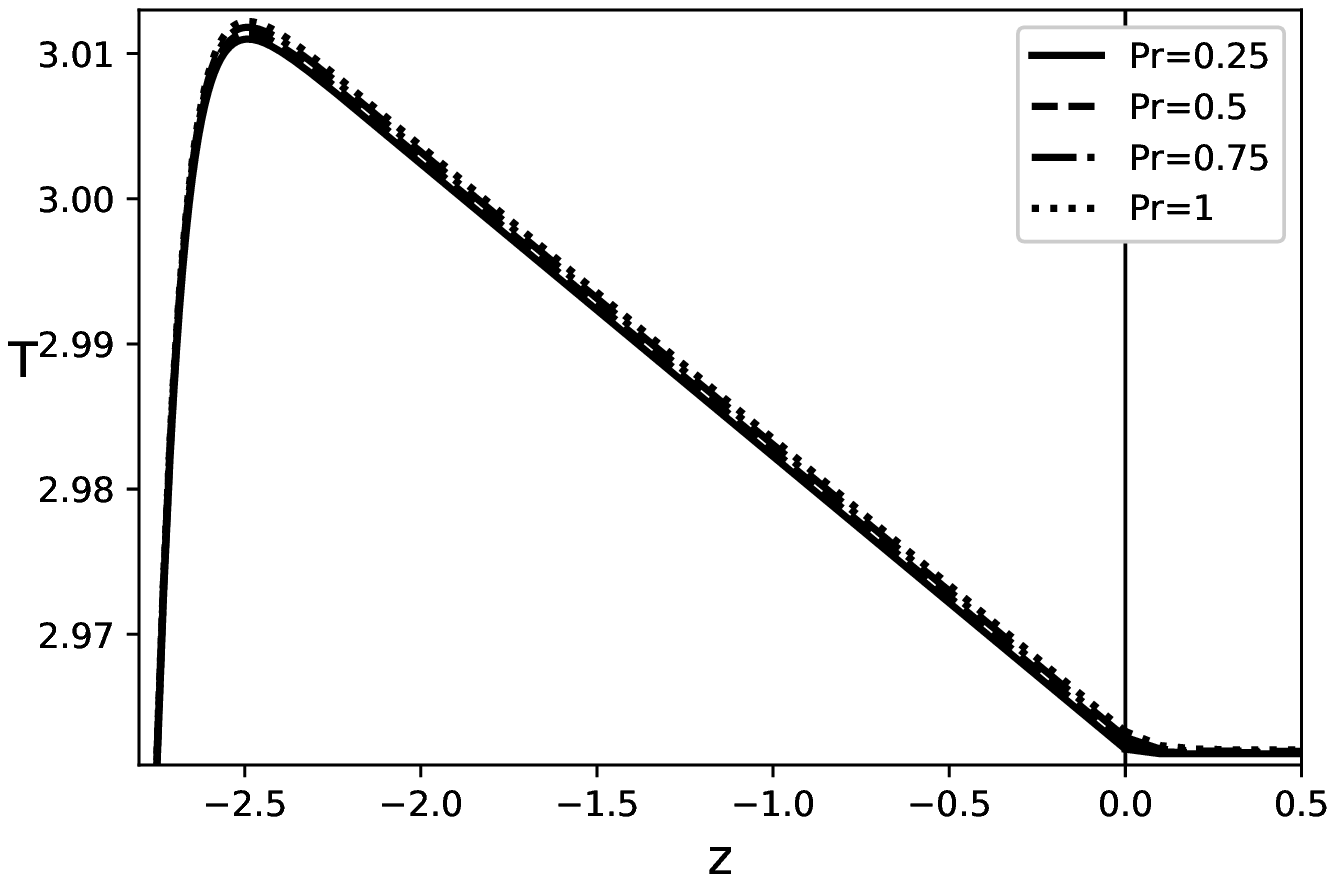}
		\end{center}
		\caption{The same as Fig. \ref{fig2} but for several values of Prandtl number with ${M\hspace{-1.5pt}a}=0.11$.}
		\label{fig4}
	\end{minipage}
\end{figure}

How the variation of $Pr$ influences the temperature profile is investigated in Figs. \ref{fig3} and \ref{fig4}.
The opposed effects are observed in positive or negative values of temperature gradient, respectively.
In the case of $d\hat{\theta}_{1}/d\eta|_{-}>0$, the increase of $Pr$ leads to the decrease of temperature as shown in Fig. \ref{fig3}.
On the other hand, for $d\hat{\theta}_{1}/d\eta|_{-}<0$, Fig. \ref{fig4} indicates that the temperature is enhanced by the viscous effect.

The effect of non-adiabaticity on the Mach number ${M\hspace{-1.5pt}a}$ is found in Table \ref{tab:value of SL}, accompanied with the value of $I$.
Clearly, the rise of temperature gradient induces the decrease of ${M\hspace{-1.5pt}a}$.
Therefore, we may consider that fast deflagration waves possess the non-adiabatic temperature profile as classically indicated by ZND Theory for one dimensional detonation wave \cite{Zel'dovich60,Neumann42,Doring43}.


\begin{table}[t]
	\caption{The values of $I$ and $d\hat{\theta}_{1}/d\eta|_{-}$, which is the temperature gradient at the burned-side edge of a flame front given in (\ref{eqn:heat_reaction_1_positive_integrate_0+}) are computed for several values of ${M\hspace{-1.5pt}a}$.
		Parameters are set as $q=2$, $N=90$, $\gamma=1.4$, ${M\hspace{-1.5pt}a}_{zero}=0.1$ and $Pr=3/4$.
	}
	\label{tab:value of SL}
	\centering
	\scalebox{0.86}{
		\begin{tabular}{ccc}
			\hline
			\rule[0pt]{0pt}{11pt}
			${M\hspace{-1.5pt}a}$ & $I$ & $d\hat{\theta}_{1}/d\eta|_{-}$\\
			\hline
			\rule[0pt]{0pt}{11pt}
			$0.12$ & $41.540$ & $-0.0257$ \\
			$0.11$ & $34.046$ & $-0.0184$ \\
			$0.1$ & $28.041$ & $-0.00835$ \\
			$0.095$ & $23.829$ & $-0.0024$ \\
			$0.09$ & $21.364$ & $0.005$ \\
			$0.085$ & $18.385$ & $0.0142$ \\
			$0.08$ & $15.399$ & $0.0264$ \\
			$0.075$ & $12.796$ & $0.042$ \\
			\hline
		\end{tabular}
	}
\end{table}

\section{CONCLUSIONS}
\label{sec:conclusions}
We have employed the method of matched asymptotic expansions with respect to $\epsilon$ ($=T_b^2/N$) under the assumption of large activation energy $N\gg1$.
By assuming that the order of Mach number $M\hspace{-1.5pt}a$ is squared root of $\epsilon$, the compression zone, whose thickness is $O({M\hspace{-1.5pt}a}^2)$, has been introduced behind a premixed flame front.
Inside of the compression zone, the effect of compressibility originates from the pressure variation due to the existence of viscosity in the Navier-Stokes equation.
The Navier-Stokes equation is solved analytically inside the compression zone to reveal the inner solutions of velocity and pressure are expressed by Lambert $W$ function.
Then, the consideration of pressure term in the heat-conduction equation leads to the non-adiabatic variation of temperature with its asymptotic solution obtained.
In the reaction zone, the analytical solution of temperature is not obtained due to the existence of reaction term.
Instead, the temperature profile is captured by solving the heat-conduction equation numerically subject to non-zero gradient condition of temperature at the burned-side edge of a flame front.

Unlike the previous works, the derivative of combination of temperature and mass fraction, or enthalpy, is no longer zero at the unburned-side edge of reaction zone.
This is because the temperature gradient at the burned-side edge of reaction zone is not restricted to zero and there is a possibility of positive or negative gradient.
The non-adiabatic feature of a flame front is observed by examining non-zero values of temperature gradient $d\hat{\theta}_{1}/d\eta|_{-}\neq0$.
If $d\hat{\theta}_{1}/d\eta|_{-}>0$, the monotonic increase of temperature is observed, which is similar with that in the adiabatic case, except for non-zero temperature gradient on the burned side of a flame front.
On the other hand, in the case of $d\hat{\theta}_{1}/d\eta|_{-}<0$, the temperature shows its non-monotonic behavior and has a single peak structure inside the reaction zone.
This temperature profile is consistent with that obtained in previous works on thermonuclear flames for small Mach numbers.
Besides, the viscosity brings the decrease of temperature for $d\hat{\theta}_{1}/d\eta|_{-}>0$ and the rise of it for $d\hat{\theta}_{1}/d\eta|_{-}<0$, though it has a slight impact.

The main assumption of our study is ${M\hspace{-1.5pt}a}=O(1/\sqrt{N})$.
Due to this specific order of Mach number, our results strongly depend on the value of non-dimensional activation energy $N$.
The removal of such a specific dependence and the study of general values of Mach number need to be treated in the future work.


\section*{Acknowledgements}
This work was supported by the Sasakawa Scientific Research Grant from The Japan Science Society (Grant Number 2022-2002).
The author is grateful to Professor Makoto Hirota for fruitful discussions and kind support.

\section*{Disclosure statement}
No potential conflict of interest was reported by the author.

\bibliographystyle{plain}
\bibliography{references}

\end{document}